\documentclass[11pt,a4paper]{article}

\usepackage{amssymb}
\usepackage{amsmath}
\usepackage{amscd}
\usepackage{latexsym}
\usepackage{epsfig}
\usepackage{xcolor}
\usepackage{fancyhdr}
\usepackage[bf,footnotesize]{caption}
\usepackage{graphicx}
\usepackage{booktabs}
\usepackage{a4wide}
\usepackage{cancel}
\usepackage{MnSymbol}
\usepackage{algpseudocode, algorithm}

\usepackage[colorlinks=true,a4paper=true,backref=false, linktocpage=true,
citecolor=blue,urlcolor=blue,linkcolor=blue,pdfpagemode=UseOutlines]{hyperref}

\hypersetup{%
  bookmarksnumbered=true,
  pdftitle = {},
  pdfsubject = {},
  pdfauthor = {U.~Wenger},
  pdfkeywords = {}
}

\graphicspath{{Figures/}}

\def\ln{\mathop{\rm ln}}                
\def\Dslash{\hbox{D}\kern-0.6em\raise0.15ex\hbox{/}} 

\def\N{{\mathbb N}}                     
\def\Z{{\mathbb Z}}                     

\def\P{{\cal P}}                     

\def\D{{\cal D}} 
\def\F{{\cal F}} 
\def\B{{\cal B}} 
\def\calN{{\cal N}}

\def\mod{\text{ mod }}               

\def\bi{\begin{itemize}}
\def\ei{\end{itemize}}
\def\be{\begin{equation}}
\def\ee{\end{equation}}
\def\beq{\begin{equation}}
\def\eeq{\end{equation}}
\def\beqs{\begin{align}}
\def\eeqs{\end{align}}

\def\beq{\begin{equation}}
\def\eeq{\end{equation}}
\def\beqs#1\eeqs{\beq\begin{split} #1 \end{split}\eeq}

\def\comment#1{}

\def\av#1{\delimiterfactor=800 \delimitershortfall=5pt \left\langle #1
  \right\rangle }

\begin{document}
\title{Solution of the sign problem in the Potts model \\at fixed fermion number}

\author{Andrei Alexandru\footnote{
Department of Physics, The George Washington University, Washington,
D.C.~20052, USA}
\footnote{Department of Physics, University of Maryland, College Park, MD 20742, USA}, 
Georg Bergner\footnote{
    Friedrich-Schiller-University, Institute of Theoretical
    Physics, Max-Wien-Platz 1, D-07743 Jena, Germany}, David Schaich, Urs Wenger
\\
Albert Einstein Center for Fundamental Physics\\
Institute for Theoretical Physics\\
University of Bern\\
 Sidlerstrasse 5\\
CH--3012 Bern, Switzerland\vspace{0.3cm}
\\
}
\date{May 1, 2018}

\maketitle

\vspace{-0.3cm}
\begin{abstract}
\noindent
We consider the heavy-dense limit of QCD at finite fermion density in
the canonical formulation and approximate it by a 3-state Potts
model. In the strong coupling limit, the model is free of the sign
problem. Away from the strong coupling, the sign problem is solved by
employing a cluster algorithm which allows to average each cluster
over the $\Z(3)$ sectors. Improved estimators for physical quantities
can be constructed by taking into account the triality of the
clusters, that is, their transformation properties with respect to
$\Z(3)$ transformations.
\end{abstract}

\section{Introduction}
At present, the non-perturbative properties of QCD at finite baryon
density can not be studied from first principles, because the theory
suffers from a fermion sign problem as soon as the baryon chemical
potential $\mu$ is nonzero. More precisely, the fluctuating sign of
the fermion determinant in the path integral at nonzero $\mu$
prevents the interpretation of the Boltzmann factor as a probability
measure and hence renders importance sampling methods such as Monte
Carlo (MC) simulations of lattice QCD inapplicable. One common
approach to the problem is to include the fluctuating sign of the
Boltzmann factor in the measured observables. While this is in
principle a valid procedure, in practice it is bound to fail because
the severe cancellations require the statistics to grow exponentially
with the space-time volume of the lattice.

Nevertheless, it turns out that in some limiting cases the fermion
sign problem is mild or even absent. One such situation concerns
lattice QCD in the so-called heavy-dense limit in the canonical
formulation at infinitely strong
coupling. In the heavy-dense limit, the fermion contributions to the
path integral are encoded in Polyakov loops associated with static
quarks and antiquarks propagating in time. In the strong coupling
limit, the Polyakov loops decouple in the effective action and the
global $\Z(3)$ symmetry of QCD at zero density is promoted to a local
one. This in turn allows to project the propagating quark states in
the canonical formulation exactly onto mesonic and baryonic states
localized on single sites. The contributions of those states to the
path integral have a phase which is zero or extremely small, hence
allowing MC calculations. Despite this happy state of affairs in the
strong coupling limit, as soon as the coupling is tuned away from this
limit, the sign problem strikes back with vengeance -- it swamps away
any coherent contribution from the colour neutral states and
consequently renders MC simulations impossible.

However, the physical picture emerging in the strong coupling limit
indicates how the relevant fermionic contributions to the path
integral, being physical and having small or even zero phases, may
look like away from infinitely strong coupling. In fact, one may
speculate that whenever the quarks and antiquarks form mesonic and
baryonic clusters, within which the Polyakov loops fluctuate in
conjuction, a coherent physical contribution should arise, very
similar to what happens in the strong coupling limit. In order to
substantiate this speculation and study the mechanism in more detail,
in this paper we simplify the problem further and consider a system in
which the gauge dynamics is replaced by that of the $\Z(3)$ Potts
model and the Polyakov loops are represented by the Potts spins $z \in
\Z(3)$ \cite{Bartholomew:1983jv,Hasenfratz:1983ce,Patel:1983sc,DeGrand:1983fk}. For
the canonical formulation of this model, where the quarks are
represented in terms of quark occupation numbers, we are able to
devise a cluster algorithm that solves the fermion sign problem
completely, in a similar way as the cluster algorithm in
\cite{Alford:2001ug} for the grand-canonical formulation with a finite
chemical potential. However, in the canonical situation described in
this paper, the solution has a clear and very intriguing physical
interpretation in terms of the quark occupation numbers, very much in
line with the physical picture that emerges for QCD in the heavy-dense
and strong coupling limit as described above.

In the cluster formulation it turns out that the only physical,
nonzero contributions to the canonical partition functions are
exclusively from clusters which contain exactly a multiple of three
quarks. The weights of these clusters are invariant under $\Z(3)$
transformations and we denote the corresponding clusters as triality-0
clusters. Clusters with one or two additional quarks are denoted as
triality-1 and triality-2 clusters, respectively, refering to the
transformation properties of their weights with respect to $\Z(3)$
transformations. Their total contributions to the canonical partition
functions exactly vanish, but they contribute to observables which
consist of nonzero-triality quark fields.

The clusters which are formed in the canonical picture allow for a
straightforward physical interpretation. The triality-0 clusters
containing a nonzero number of quarks represent baryon or multi-baryon
states. While the quarks can move freely within the cluster, of course
always respecting the Pauli exclusion principle, they are nevertheless
confined within the clusters. It is therefore obvious to interpret
these clusters as baryon bags. Empty clusters on the other hand
represent vacuum fluctuations of the underlying gauge degrees of
freedom. In case we also allow antiquarks in the system, the
triality-0 clusters may also contain mesons and represent all kinds of
multi-baryon-multi-meson bags.

This physical picture is also useful to interpret the transition from
the confined phase to the deconfined one. At low temperature and
density, the clusters tend to be very small and form a dilute gas of
hadrons, representing the system in the confined phase. When the
temperature is increased, the clusters grow larger and start to
percolate, at which point the system undergoes a first order phase
transition into the deconfined phase. At even higher temperatures, one
single cluster essentially fills the whole volume and the quarks can
hence move freely within the whole volume.

Before moving on to work this picture out in full detail, it is useful
to discuss the symmetry properties of the $\Z(3)$ Potts model at
finite density. While the model at zero fermion density is invariant
under global $\Z(3)$-symmetry transformations, this is no longer true
at finite density. In that case the introduction of the chemical
potential breaks the $\Z(3)$\,symmetry explicitly.  In contrast, in
the canonical formulation the contributions from the static charges
have definite transformation properties under $\Z(3)$ transformations,
such that the contributions vanish unless the number of charges is a
multiple of 3. As a consequence, the non-vanishing contributions are
manifestly invariant under global $\Z(3)$-symmetry transformations.

The paper is organised as follows. Section \ref{sec:Potts model}
contains the definition of the $\Z(3)$ Potts model and its canonical
formulation. In section \ref{sec:bond formulation} we introduce the
bond formulation and describe the cluster algorithm for the canonical
formulation, while in section \ref{sec:improved estimators} we
construct improved estimators for various physical quantities such as
the quark-antiquark, the quark-quark, the antiquark-antiquark and the
3-quark correlator, as well as the free energy of a single quark and
of a single antiquark. A practical algorithm to sample the various
sectors in configuration space based on the calculation of the
multiplicities of quark configurations with subsequent reweighting is
presented in section \ref{sec:multiplicities}. Next, in section
\ref{sec:Severity of the sign problem} we discuss the severity of the
sign problem, while in section \ref{sec:results} we present a
selection of simulation results, including an illustration of how the
expected phase transition at very low fermion density manifests itself
in the canonical setup. Finally, section \ref{sec:conclusions}
contains our conclusions, while some technical details are collected
in the appendices.

\section{The 3-dimensional $\Z(3)$ Potts model}
\label{sec:Potts model}
In order to introduce the 3-dimensional $\Z(3)$ Potts model we follow
\cite{Alford:2001ug} and approximate the grand-canonical partition
function of QCD in the heavy-dense limit by
\begin{equation}
\label{eq:grand-canonical partition function}
Z_\text{GC}(h) = \int \D z \, \exp(-S[z] + h \sum_x z_x)
\end{equation}
where the Potts spins $z_x \in \Z(3)$ are defined on discrete lattice
points $x$ and stand for the Polyakov loop variables generated by the
infinitely heavy quarks. The parameter $h=(2\kappa e^\mu)^\beta$
refers to the hopping parameter $\kappa$, the chemical potential $\mu$
and the inverse temperature $\beta$ of the original theory. The
precise form of the action can in principle be derived from QCD by
integrating out all degrees of freedom except for the $\Z(3)$ phase of
the Polyakov loop. In the Potts model $h$ plays the role of an
external (magnetic) field which couples to the Potts spins and hence
explicitly breaks the $\Z(3)$ symmetry present for a
$\Z(3)$-invariant action $S[z]$ and $h=0$.

In the canonical formulation the partition function for $N_Q$ quarks
can be written in the generic form
\begin{equation}
\label{eq:canonical partition function}
Z_\text{C}(N_Q) = \sum_{\{n\}} \int \D z \, \exp(-S[z]) \cdot \prod_x g[z_x,n_x] \, .
\end{equation}
Here the quark occupation numbers $n_x \in \{0,\ldots,n_\text{max}\}$
are collected in the quark configuration $n=\{n_x\}$ and the sum is
over all quark configurations which fulfill $|n| \equiv \sum_x n_x =
N_Q$. The finite maximal number of quarks $n_\text{max}$ on each site
implements the Pauli exclusion principle. The fermionic weights
$g[z_x,n_x]$ can in principle be derived from QCD and in general also
involve the anti-Polyakov loops $z_x^*$ and corresponding occupation
numbers $\bar n_x$ for the antiquarks. Here we replace the fermionic
weights by
\begin{equation}
\label{eq:simple fermionic weight}
g[z,n] = z^n \, .
\end{equation}
For this case, the connection to the grand-canonical partition
function in eq.(\ref{eq:grand-canonical partition function}) can be
made explicit and it can be shown that the correspondence is exact for
$h\ll 1$, i.e.~at low densities. We refer to Appendix
\ref{app:connection grand-canonical/canonical} for further details.

For the action $S[z]$ representing the pure gluon action we choose the
standard nearest-neighbour Potts model interaction
\begin{equation}
S[z] = - \gamma \sum_{\langle x y\rangle} \delta_{z_x,z_y}
\end{equation}
where the sum is over all nearest-neighbour lattice sites. Note that
the total action including the fermionic weights is manifestly complex
and suffers from a sign problem just as in the grand-canonical
formulation. However, the properties of the action under global
$\Z(3)$ transformations ensure that
\begin{equation}
Z_{N_Q \neq 0 \mod 3} = 0 \, .
\end{equation}
In the limit $\gamma \rightarrow 0$, corresponding to the strong
coupling limit in QCD, the Potts spins fluctuate independently on each
site and the global $\Z(3)$ transformation is promoted to a local one
which hence enforces
\begin{equation}
n_x = 0 \mod 3,  \quad \forall x \qquad \text{('strong coupling limit' $\gamma \rightarrow 0$)}.
\end{equation}
The fermionic weights then become trivial and the partition function
simply counts the number of allowed quark configurations,
\begin{equation}
\lim_{\gamma \rightarrow 0} Z_{N_Q} = \sum_{\{n\}}\int \D z = 3^V \cdot G(N_Q/3) \, . 
\end{equation}
In case the quark occupation number is restricted to $n_x \le
n_\text{max} = 3$, the number of allowed quark configurations yields $G(N_Q/3) =
{{V}\choose{N_Q/3}}$.

\section{Bond formulation and cluster algorithm}
\label{sec:bond formulation} 
For finite gauge coupling the canonical partition function reads
\begin{equation}
Z_{N_Q} = \sum_{\{n\}}\int \D z \exp(\gamma \sum_{\langle xy \rangle} \delta_{z_x,z_y}) \prod_x z_x^{n_x}
\end{equation}
and with the help of
\begin{equation}
e^{\gamma \cdot \delta_{z_x,z_y}} = \sum_{b_{xy}=0}^1 \left( \delta_{z_x,z_y} \delta_{b_{xy},1} \left(e^\gamma - 1\right) + \delta_{b_{xy},0} \right)
\end{equation}
it can be written as
\begin{equation}
\label{eq:Z_{N_Q} with n, b z}
Z_{N_Q} = \sum_{\{n\}}\sum_{\{b\}}\int \D z \prod_{\langle xy \rangle}\left( \delta_{z_x,z_y} \delta_{b_{xy},1} \left(e^\gamma - 1\right) + \delta_{b_{xy},0} \right) \prod_x z_x^{n_x}
\end{equation}
where the sum is over all bond configurations $b=\{b_{xy}\}$.  For
future reference we introduce a double bracket notation to abbreviate
the sum over $\{b\}$, $\{n\}$ and ${\cal D}z$ weighted by the bond
factors. For a generic quark number $N_Q = \sum_x n_x$ and a generic
function $f(\{z_x\})$ we define
\begin{equation}
\llangle
\prod_x z_x^{n_x} \cdot f(\{z_x\})\rrangle_{N_Q} = \sum_{\{n\}}\sum_{\{b\}}\int \D z
\prod_{\langle xy \rangle}\left( \delta_{z_x,z_y} \delta_{b_{xy},1}
  \left(e^\gamma - 1\right) + \delta_{b_{xy},0} \right) \prod_x
z^{n_x} \cdot f(\{z_x\}) \, ,
\end{equation}
i.e., eq.(\ref{eq:Z_{N_Q} with n, b z}) becomes
\begin{equation}
Z_{N_Q} = \llangle \prod_x z_x^{n_x}  \rrangle_{N_Q}
\end{equation}

Let us now consider for a moment the case $N_Q=0$ when there are no
fermionic contributions through the Polyakov loops. From the
expression above we read off that for a given spin configuration a
bond $b_{xy}$ is occupied with probability $p(b_{xy}=1) =
(1-e^{-\gamma})$ if the two neighbouring spins $z_x$ and $z_y$ are
aligned, else it is empty ($b_{xy}=0$). The weight of such a bond
configuration is then simply $W(\{b\}) = (e^\gamma-1)^{N_b}$, where
$N_b=\sum_{\langle xy\rangle} b_{xy}$ is the total number of occupied
bonds, and it is independent of the orientation of the spins within
the connected clusters of bonds.  Hence, all spins within a cluster
can be flipped into any of the three states independently of the spins
in the other clusters. Up to here, this is just the well-known
Swendsen-Wang cluster reformulation \cite{Swendsen:1987ce} of the
Potts model. One can now integrate out the spins for a given bond
configuration by summing over all possible combinations of cluster
orientations, yielding
\begin{equation}
\label{eq:Swendsen-Wang cluster partition function}
Z_{N_Q=0} =\llangle 1 \rrangle_{N_Q=0} = \sum_{\{b\}} (e^\gamma -1)^{N_b} \cdot 3^{N_C}
\end{equation}
where $N_C$ is the number of clusters for the given bond
configuration. This is just the random cluster model representation of
the Potts model as derived by Fortuin and Kasteleyn in
\cite{Fortuin:1971dw}. Note that the integration over the spins is
possible because the bond variables decouple the spin orientations of
the clusters from each other, so that the integration over the spin
degrees of freedom can be factorized over the clusters. This
factorization is in fact the basis for the solution of the sign
problem when the fermionic contributions are included.

In order to simulate the system according to the weights given in
eq.(\ref{eq:Swendsen-Wang cluster partition function}), the bonds can
be updated with a local algorithm as follows. A bond whose value does
not change the number of clusters is activated with probability
$p=1-e^{-\gamma}$. A bond which connects two clusters $C_1$ and $C_2$,
which would otherwise be separate, is a bridging bond and therefore
called a {\it bridge} in short. Activating a bridge decreases the
number of clusters $N_C$ by one, and it is therefore done with a
probability
\begin{equation}
\label{eq:connect probability}
 p(C_1,C_2 \rightarrow C_1\cup C_2)=\frac{e^\gamma - 1}{e^\gamma + 2}=\frac{1-e^{-\gamma}}{1+2 e^{-\gamma}} \, ,
\end{equation}
while the bridge is deactivated  with the probability  
\begin{equation}
\label{eq:breakup probability}
 p(C_1\cup C_2 \rightarrow C_1, C_2)=\frac{3}{e^\gamma + 2} =\frac{3 e^{-\gamma}}{1+2 e^{-\gamma}}\, .
\end{equation}
Hence, the update of a bond requires one to check whether the bond
under consideration is a bridge or not. This is a well known,
difficult problem in computational complexity theory and goes under
the name {\it fully-dynamic connectivity problem}. For further details
and our strategy to deal with the problem we refer to Appendix
\ref{app:fully-dynamic connectivity}.

For $N_Q>0$ we can include the fermionic contribution in our
considerations by constructing an improved estimator for it. For an
individual configuration the contribution $\prod_x z^{n_x}$ is in
general complex, but its average over the subensemble of the $3^{N_C}$
configurations related by the $\Z(3)$ transformations of the
individual clusters is rather simple. In the following we denote this
subensemble average by $\langle \, \cdot \, \rangle_{3^{N_C}}$. We
first observe that the total weight can be factorized into individual
cluster weights $W_0(C)$,
\begin{equation}
  \langle \prod_x z_x^{n_x} \rangle _{3^{N_C}} = \langle \prod_C
  \prod_{x\in C} z_x^{n_x}\rangle _{3^{N_C}} = \prod_C \langle
  \prod_{x\in C} z_x^{n_x}\rangle_{3} = \prod_C W_0(C) \, ,
\end{equation}
where by $\langle \, \cdot \, \rangle_{3}$ we denote the average over
the three $\Z(3)$ orientations of a given cluster. This average can be
calculated due to the fact that all spins in the cluster are aligned,
\begin{equation}
W_0(C) = \langle\prod_{x\in C} z^{n_x}\rangle_{3} = \langle z^{ \sum_{x\in C} n_x}\rangle_{3} = \left\{ 
\begin{array}{ll}
1 & \text{if } \sum_{x\in C} n_x =0 \mod 3 , \\
0 & \text{else}.
\end{array}
\right.
\end{equation}
Denoting the fermion number content of each cluster modulo 3 by
\begin{equation}
n_C= \sum_{x\in C} n_x \mod 3 \, ,
\end{equation}
the cluster weight simply becomes $W_0(C) = \delta_{n_C,0}$. In the
following we refer to $n_C$ as the triality of the cluster
$C$. Consequently, all bond clusters can be classified according to
their fermion content $n_C$, i.e.~their triality. Clusters with
nonzero triality $n_C=1$ or 2 have a vanishing contribution to the
partition function,\footnote{In other contexts such clusters are
  sometimes called meron clusters.}  while clusters with zero triality
$n_C=0$ contribute positively with weight 1. Hence, the partition
function can now be written as
\begin{equation}\label{eq:Z_NQ}
Z_{N_Q} = \llangle \prod_x z_x^{n_x}\rrangle_{N_Q}  = \sum_{\{n\}}\sum_{\{b\}} (e^\gamma -1)^{N_b} \cdot 3^{N_C}
\cdot \prod_C \delta_{n_C,0}
\end{equation}
where the sums are with respect to all bond and occupation number
configurations (with $\sum_x n_x = N_Q$). The fermionic contribution
$\prod_x z_x^{n_x}$ simply projects onto the sector of configurations
containing only triality-0 clusters. Note that although the
contributions to the partition function are now all positive, the
expectation value of the fermionic contribution with respect to the
full, unrestricted ensemble with fixed $N_Q$,
\begin{equation}
\label{eq:complex phase vev}
 \langle \prod_C \delta_{n_C,0} \rangle_{N_Q} = \frac{\llangle \prod_x z_x^{n_x} \rrangle_{N_Q}}{\llangle 1 \rrangle_{N_Q}} \, ,
\end{equation}
can still be exponentially small, because the probability to find
configurations without nonzero-triality clusters among all possible
cluster and fermion number configurations can be exponentially small
in the volume. In fact, it is easy to see that eq.(\ref{eq:complex
  phase vev}) is just the expectation value of the complex phase in
the modified real action ensemble and represents a measure for the
severity of the sign problem. We refer to Sec.~\ref{sec:Severity of
  the sign problem} for a more detailed discussion.

Nevertheless, the partition function can in principle be updated by
directly operating on the bond variables and the fermion occupation
numbers. Updates need to ensure that the system stays in the
configuration space containing only triality-0 clusters. The bond
update can be implemented as before with the cluster break-up
probability in eq.(\ref{eq:breakup probability}) modified to
\begin{equation}
\label{eq:zero-triality breakup probability}
 p(C_1\cup C_2 \rightarrow C_1, C_2)=\frac{3}{e^\gamma + 2} \cdot \delta_{n_{C_1},0} \cdot \delta_{n_{C_2},0}=\frac{3 e^{-\gamma}}{1+2 e^{-\gamma}} \cdot \delta_{n_{C_1},0} \cdot \delta_{n_{C_2},0}\, ,
\end{equation}
where the $\delta$-functions make sure that no clusters with nonzero
triality are generated. Note that this step fulfills detailed balance
since the situation involving the reverse step, i.e.~connecting two
clusters with at least one having nonzero triality, can never occur.
Once the bond configuration is updated, the fermion occupation numbers
can be updated locally by proposing to shift a fermion from site $x$
to one of its nearest-neighbour sites $y$. The shift is accepted with
probability 1 as long as the fermion remains in the same cluster.
Hops to sites which are already saturated are forbidden, as well as
hops into a different cluster, since these would leave behind two
clusters with nonzero triality. Hence, for two neighbouring sites $x$
and $y$ we have
\begin{equation}\label{eq:no merons fermion hop}
p(n_x, n_y \rightarrow n_x-1,n_y+1) = (1-\delta_{n_x,0})(1-\delta_{n_y,n_\text{max}})\cdot \delta_{C_x,C_y}
\end{equation}
where $C_{x}$ and $C_y$ denote the clusters containing the sites $x$
and $y$, respectively. Additional update steps can of course be
envisaged and might be favourable with respect to efficiency of the
algorithm. For example, one could let three fermions, i.e.~a baryon,
hop together. Since this update does not change the triality of the
clusters, the hop is also allowed across the clusters.

In a sense, the algorithm realizes a kind of a bag model in which the
fermions can move freely within bags defined by the individual
clusters, i.e.~the fermions are confined to the clusters, but
deconfined within the clusters. In the confined phase we expect the
clusters to be small, so the quarks within a cluster are confined
within the small regions of the clusters. In the limit $\gamma
\rightarrow 0$ no bonds are allowed and the clusters consist of single
sites only, so the quarks are bound into baryons confined to single
sites, just as it happens for QCD in the strong coupling
limit. Towards the deconfined phase the clusters proliferate, so the
quarks can move freely within large regions and hence form baryons of
extended size. In the deconfined phase, the clusters percolate the
whole volume, so the quarks can move unrestricted throughout the whole
volume and the baryon loses its meaning as a bound state of three
quarks. Hence, what we have here is a set of degrees of freedom which
captures and describes the physics of the underlying system very
naturally.

\section{Improved estimators for physical quantities}
\label{sec:improved estimators}
The reformulation in terms of bond variables and fermion occupation
numbers allows to construct improved estimators for physical
observables. First we note that the expectation values of observables
with nonzero triality vanish in the ensembles with $N_Q=0 \mod
3$. Hence, for the partition function of a single (anti-)Polyakov loop
in the restricted ensemble without nonzero-triality clusters we have
\begin{equation}
\llangle z_x \prod_y z_y^{n_y} \rrangle_{N_Q=0\mod 3} = \llangle z_x^*
\prod_y z_y^{n_y}\rrangle_{N_Q=0\mod 3} = 0 \, ,
\end{equation}
which reflects the fact that the free energy of a single quark or
antiquark is infinite in the background of integer baryon number,
$F_{q}|_{N_Q/3 \in \N} =F_{\bar q}|_{N_Q/3 \in \N}\rightarrow
\infty$ due to Gauss' law. The same holds true for the quark-quark or
antiquark-antiquark correlation functions in the background of integer
baryon number. In contrast, a single quark or antiquark can exist as a
probe in a background with non-integer baryon number. Indeed, the
partition function including a single quark or antiquark as a source
picks up contributions exclusively from the sectors of configurations
which contain nonzero-triality clusters,\footnote{Here and in the
  following we denote the inverse temperature by $\beta$, as usual.}
\beqs
\llangle z_x \prod_y z_y^{n_y}\rrangle_{N_Q=2\mod 3} &= \exp(-\beta F_{q_x} |_{N_Q=2\mod 3})\neq 0, \\
\llangle z_x^* \prod_y z_y^{n_y}\rrangle_{N_Q=1\mod 3} &= \exp(-\beta
F_{\bar q_x} |_{N_Q=1\mod 3}) \neq 0 \, ,
\eeqs
and similarly for the partition functions including a quark-quark or
an antiquark-antiquark correlator.

To be more precise, let us construct an improved estimator for the
partition function of a single quark at position $x$ in the background
of $N_Q=2\mod 3$ quarks.  For a given cluster configuration and
fermion number configuration (with $\sum_y n_y = N_Q-1$ and $N_Q=0\mod
3$) we denote by $C_x$ the cluster containing the site $x$ and
calculate the subensemble average according to
\begin{align}
  \langle z_x \prod_y z_y^{n_y} \rangle _{3^{N_C}} 
&= \langle z_x \prod_{y\in C_x} z_y^{n_y} \prod_{C\neq C_x}
\prod_{y\in C} z_y^{n_y}\rangle _{3^{N_C}} \nonumber \\
&= \langle \prod_{y\in C_x} z_y^{n_y+\delta_{x,y}}\rangle_{3}   \prod_{C\neq C_x} \langle \prod_{y\in C} z_y^{n_y}\rangle_{3} \\ 
&= W_2(C_x) \cdot \prod_{C\neq C_x} W_0(C) \, .   \nonumber 
\end{align}
As before the weights $W_0(C) = \delta_{n_C,0}$ project onto the 
triality-0 clusters, while the weight $W_2(C_x)$ is 
\begin{equation}
\label{eq:def W2}
W_2(C_x) = \langle \prod_{y\in C_x} z^{n_y+\delta_{x,y}}\rangle_{3} = \left\{  
\begin{array}{ll}
1 & \text{if } \sum_{y\in C_x} n_y =2 \mod 3 \\
0 & \text{else}
\end{array}
\right\} = \delta_{{n_{C_x}},2} 
\end{equation}
i.e.~it is nonzero only if $C_x$ is a cluster with triality
$n_{C_x}=2$. Hence, after integrating out the spin variables, the
observable for a single quark at position $x$ in the background of
$N_Q=2\mod 3$ quarks is just
\begin{equation}
z_x \, \rightarrow \, \delta_{n_{C_x},2} \cdot \prod_{C\neq C_x}\delta_{n_C,0} \, .
\end{equation}
The observable simply counts the number of configurations with all
clusters having triality $n_C=0$ except the one containing the site
$x$ which has triality $n_{C_x}=2$. Similarly, the observable for an
antiquark in the background of $N_Q=1\mod 3$ quarks can be written as
\begin{equation}
z_x^* \, \rightarrow \, \delta_{n_{C_x},1} \cdot \prod_{C\neq
  C_x}\delta_{n_C,0}
\label{eq:improved estimator for zbar}
\end{equation}
which counts the number of configurations with all clusters having
triality $n_C=0$ except the cluster $C_x$ with triality
$n_{C_x}=1$. The first $\delta$-function in eq.(\ref{eq:improved
  estimator for zbar}) is just the weight $W_1(C_x)$ defined in
analogy to eq.(\ref{eq:def W2}) by
\begin{equation}
W_1(C_x) = \langle \prod_{y\in C_x} z_y^{n_y-\delta_{x,y}}\rangle_{3} = \left\{  
\begin{array}{ll}
1 & \text{if } \sum_{y\in C_x} n_y =1 \mod 3 \\
0 & \text{else}
\end{array}
\right\} = \delta_{{n_{C_x}},1} \, .
\end{equation}

Of course, while the absolute free energy can not be measured directly
in a Monte Carlo simulation, the free energy difference can be
determined from appropriate ratios of partition functions. The free
energy difference between $N_Q=0\mod 3$ quarks and a quark in the
background of $N_Q -1$ quarks for example can be calculated from
\begin{equation}
\label{eq:Delta F_q}
\exp(-\beta \Delta F_{q_x})|_{N_Q} = \exp(-\beta (F_{q_x}|_{N_Q-1} - F|_{N_Q})) =
\frac{\llangle z_x \prod_{y}z_y^{n_y} \rrangle_{N_Q-1}}{\llangle
  \prod_{y}z_y^{n_y} \rrangle_{N_Q}} \equiv \langle z_x\rangle_{N_Q}
\end{equation}
corresponding to $\langle z_x\rangle_{\rho}$ in the thermodynamic
limit $V\rightarrow \infty$ with $\rho = N_Q/V$ fixed.  The crucial
step to simulate this system with the spins already integrated out is
the observation that the bond cluster configurations contributing to
the numerator and the denominator are the same -- the only difference
is in the fermion numbers and possible fermion configurations. Hence,
we can measure the ratio by simulating both the sector where all
clusters have zero triality $n_C =0$ and the sector with one quark
less and exactly one cluster containing the site $x$ with triality
$n_{C_x}=2$.
This can be achieved by setting up a Monte Carlo process which probes
the configuration space of clusters and quark configurations with
$N_Q$ and $N_Q-1$, the latter with exactly one triality-2 cluster
connected to site $x$. More precisely, we add two update steps which
try to remove or add a quark from a random site $y$ depending on
whether the system is in the sector with $N_Q=0 \mod 3$ or $N_Q-1$
quarks. The removal is only accepted when the site $y$ is in the same
cluster as $x$, so for a random site $y$ we have
\begin{equation}
p(n_y \rightarrow n_y-1) = (1 - \delta_{n_y,0}) \cdot \delta_{C_x,C_y} \cdot
\delta_{N_q, 0 \mod 3}\, ,
\end{equation}
while the balancing acceptance probability for adding a quark is
\begin{equation}
p(n_y \rightarrow n_y+1) = (1 - \delta_{n_y,n_\text{max}}) \cdot
\delta_{C_x,C_y} \cdot \delta_{n_{C_y,2}}\, .
\end{equation}
Moreover, in the sector $N_Q-1$ the triality-2 cluster can only be a
broken in two if the triality-2 cluster remains connected to site
$x$, so the cluster break-up probability is given by
\begin{equation}
\label{eq:triality-zero breakup probability 2}
p(C_1\cup C_2 \rightarrow C_1, C_2)=\frac{3}{e^\gamma + 2} \cdot 
\left\{
\begin{array}{ll}
\delta_{n_{C_1},2} \cdot \delta_{n_{C_2},0} \cdot \delta_{C_1,C_x} &
\text{if } x\in C_1\cup C_2 ,\\
\delta_{n_{C_1},0} \cdot \delta_{n_{C_2},0} & \text{if } x\notin C_1
\cup C_2 \, . 
\end{array}
\right.
\end{equation}
We can of course make use of translational invariance of $F_{q_x}$ and
collect statistics for each site in the triality-2 cluster. One just
has to make sure that in the update steps above at most one
triality-2 cluster is present in the system. Denoting by $N_1$ and
$N_2$ the number of clusters with triality 1 and 2, respectively, this
means that the allowed configuration space is extended to include
configurations with $(N_1, N_2)=(0,0)$ {\it and} $(0,1)$
nonzero-triality clusters.

It is straightforward to construct similar update steps for simulating
both the sector where all clusters have zero triality $n_C=0$ and the
sector with one additional quark and hence exactly one cluster with
triality $n_C=1$. The allowed configuration space is then extended to
configurations with $(N_1, N_2)=(0,0)$ and $(1,0)$ nonzero-triality
clusters. The relative occurrence of configurations in the two sectors
eventually yields the free energy difference between $N_Q=0\mod 3$
quarks and a single antiquark in the background of $N_Q+1$ quarks,
\begin{equation}
\label{eq:Delta F_qbar}
\exp(-\beta \Delta F_{\bar q_x})|_{N_Q} = \exp(-\beta (F_{\bar q_x}|_{N_Q+1} - F|_{N_Q})) =
\frac{\llangle z_x^* \prod_{y}z_y^{n_y} \rrangle_{N_Q+1}}{\llangle
  \prod_{y}z_y^{n_y} \rrangle_{N_Q}} \equiv \langle z_x^*\rangle_{N_Q}
\end{equation}
corresponding to $\langle z_x^*\rangle_{\rho}$ in the thermodynamic
limit $V\rightarrow \infty$ with $\rho = N_Q/V$ fixed.

It is noteworthy that the probabilities for the transitions between
the sectors $(1,0) \rightarrow (0,0)$ and $(0,0) \rightarrow (0,1)$
are suppressed by $1/V$ because the numbers of configurations in the
corresponding configuration spaces scale accordingly. As a
consequence, the ratios in eq.(\ref{eq:Delta F_q}) and (\ref{eq:Delta
  F_qbar}) are more and more difficult to measure accurately in
numerical simulations towards the thermodynamic limit.  This can be
alleviated by artificially enhancing and suppressing the transition
probabilities between the sectors followed by an appropriate
a-posteriori reweighting.  A more drastic approach is the following.
Since the bond cluster configurations contributing to the numerator
and the denominator are the same up to the difference in the fermion
numbers, we can measure the ratios by simulating only in the sector
$(N_1,N_2)=(0,0)$ and reweighting to the fermion configurations with
$(1,0)$ and $(0,1)$ using the corresponding multiplicities. We refer
to Sec.~\ref{sec:multiplicities} for the detailed discussion of this
alternative approach.

Observables with nonvanishing expectation values in integer baryon
number sectors are for example the quark-antiquark correlator $\langle
z_x z_y^*\rangle_{N_Q=0\mod 3}$ and the 3-point quark correlator
$\langle z_x z_y z_w\rangle_{N_Q=0\mod 3}$. Let us first consider the
improved estimator for the quark-antiquark correlator. For a given
fermion number configuration with $\sum_x n_x = 0\mod 3$ and fixed
cluster configuration we calculate the subensemble average while
keeping track whether the sites $x$ and $y$ belong to the same
cluster,
\begin{align}
\langle z_x z_y^* \prod_w z_w^{n_w} \rangle_{3^{N_C}} 
&= \delta_{C_x,C_y} \cdot \langle z_x z_y^* \prod_w z_w^{n_w} \rangle_{3^{N_C}}
+ (1-\delta_{C_x,C_y})\cdot \langle z_x z_y^* \prod_w z_w^{n_w}
\rangle_{3^{N_C}} \nonumber \\
&=  \delta_{C_x,C_y} \cdot \langle \prod_{w\in C_x}
z_w^{n_w+\delta_{w,x}-\delta_{w,y}} \rangle_3 \prod_{C\neq C_x}
\langle \prod_{w\in C} z_w^{n_w}\rangle_{3} \nonumber \\
&{} \quad +  (1-\delta_{C_x,C_y})\cdot \langle \prod_{w\in C_x}
z_w^{n_w+\delta_{w,x}} \rangle_3 \cdot  \langle \prod_{w\in C_y}
z_w^{n_w-\delta_{w,y}} \rangle_3  \prod_{{C \neq C_x}\atop{C\neq C_y}}
\langle \prod_{w\in C} z_w^{n_w}\rangle_{3}\nonumber \\
&= \delta_{C_x,C_y} \cdot \prod_{C} W_0(C) + (1-\delta_{C_x,C_y}) \cdot  W_2(C_x) \cdot  W_1(C_y) \prod_{{C \neq C_x}\atop{C\neq C_y}} W_0(C)
\end{align}
where we have used that $z z^* = 1$. Expressing the weights of the
clusters explicitly in terms of $\delta$-functions, we can write the
improved estimator for the correlator $z_x z_y^*$ as
\begin{equation}
z_x z_y^* \, \rightarrow \,
\delta_{C_x,C_y} \cdot \prod_{C} \delta_{n_C,0} +
(1-\delta_{C_x,C_y}) \cdot \delta_{n_{C_x},2}\cdot  \delta_{n_{C_y},1}
\prod_{{C \neq C_x}\atop{C\neq C_y}} \delta_{n_c,0} \,.
\label{eq:improved estimator z zbar}
\end{equation}
Hence, the quark-antiquark correlator receives contributions both from
the zero triality sector and from the sector with exactly one cluster
with triality $n_{C_x}=2$ and one cluster with triality
$n_{C_y}=1$.\footnote{For $N_Q=0$ there can be no hops or breakups of
  a zero triality cluster into nonzero triality ones, hence the
  correlator gets contributions only from $\delta_{C_x,C_y}$.}
Therefore, in a numerical simulation we allow the breakup of a zero
triality cluster with $\sum_{x\in C}n_x>0$ into the two required
nonzero triality clusters, and the probability in
eq.(\ref{eq:zero-triality breakup probability}) to remove a bridge is
modified accordingly,
\begin{equation}
p(C_1\cup C_2 \rightarrow C_1, C_2)=\frac{3}{e^\gamma + 2} \cdot 
\left\{
\begin{array}{ll}
\delta_{n_{C_1},1} \cdot \delta_{n_{C_2},2} & \text{if } y\in C_1 \land x\in C_2 ,\\
\delta_{n_{C_1},0} \cdot \delta_{n_{C_2},0} & \text{else}.
\end{array}
\right.
\end{equation}
In addition, we allow a fermion to hop from the cluster $C_x$ with
zero triality into the zero triality one $C_y$, thereby generating the
two required clusters with nonzero triality. Hence, the probability in eq.(\ref{eq:no merons fermion hop}) for a fermion to hop from site $v$ to site $w$ is modified according to
\begin{multline}
\label{eq:two merons fermion hop}
p(n_v, n_w \rightarrow n_v-1,n_w+1) \\
= 
(1-\delta_{n_v,0})(1-\delta_{n_w,n_\text{max}}) \cdot
\left\{\delta_{C_v,C_w} +  (1-\delta_{C_v,C_w}) \delta_{C_v,C_x}
  \delta_{C_w,C_y} \right\} \, .
\end{multline}
In this way we efficiently probe both the sectors which contribute to
the correlator, and we obtain the quark-antiquark potential in the
background of $N_Q$ quarks as
\begin{equation}
\exp(-\beta V_{q\bar q}(x-y))|_{N_Q} 
= \llangle z_x z_y^* \prod_{w}
z_w^{n_w}\rrangle_{N_Q}/\llangle \prod_{w}
z_w^{n_w}\rrangle_{N_Q} = \langle z_x z_y^* \rangle_{N_Q}\, .
\label{eq:quark-antiquark potential}
\end{equation}
Of course one can generalize the above steps in such a way that the
allowed configuration space is extended to include configurations with
just $(N_1,N_2)=(0,0)$ and $(1,1)$ nonzero-triality clusters
independent of $x$ and $y$. By making use of translational invariance,
one can then collect statistics for all possible distances $x-y$ on a
given cluster and fermion configuration. However, since the bond
cluster configurations contributing to the numerator and the
denominator are the same up to the difference in the fermion numbers,
we can measure the ratio as in the case of $\langle z_x\rangle$,
namely by simulating in the sector containing triality-0 clusters only
and reweighting to the fermion configurations with $n_{C_x} =2$ and
$n_{C_y}=1$. We refer again to Sec.~\ref{sec:multiplicities} for a
more detailed discussion of this approach.

The improved estimators for the quark-quark and antiquark-antiquark
correlators can be constructed in an analogous way. Here we only note
that the quark-quark correlator is defined exclusively in the
background of $N_Q -2 = 0 \mod 3$ quarks, allowing only configurations
with $(N_1, N_2) = (1,0)$ and $(0,2)$ nonzero-triality clusters. In
contrast, the antiquark-antiquark correlator is defined only in the
background of $N_Q +2 = 0 \mod 3$ quarks, requiring exclusively
configurations with $(N_1, N_2) = (0,1)$ and $(2,0)$ nonzero-triality
clusters. Consequently, the physical quantity of interest is the free
energy difference, or potential energy, between $N_Q = 0 \mod
3$ quarks and  the two (anti)quarks
in the corresponding background,
\beqs
\exp(-\beta V_{q q}(x-y))|_{N_Q} 
&= \llangle z_x z_y \prod_{w}
z_w^{n_w}\rrangle_{N_Q-2}/\llangle \prod_{w}
z_w^{n_w}\rrangle_{N_Q} \equiv \langle z_x z_y \rangle_{N_Q}\, ,
\\
\exp(-\beta V_{\bar q\bar q}(x-y))|_{N_Q} 
&= \llangle z_x^* z_y^* \prod_{w}
z_w^{n_w}\rrangle_{N_Q+2}/\llangle \prod_{w}
z_w^{n_w}\rrangle_{N_Q} \equiv\langle z_x^* z_y^* \rangle_{N_Q} \, .
\label{eq:antiquark-antiquark potential}
\eeqs

Finally, let us now consider the expectation value of the 3-point
quark correlator $\langle z_x z_y z_w\rangle$ in the background of
${N_Q=0\mod 3}$ quarks. The derivation of the corresponding improved
estimator is analogous to the derivations before. When calculating the
subensemble average of the 3-point correlator $z_x z_y z_w$ including
the canonical fermion weight $\prod_v z_v^{n_v}$ one has to keep track
of which of the three spins belong to the same cluster. They can all
belong to the same cluster, or to two or three different ones. The
latter two possibilities generate contributions in the sectors with
exactly two or three clusters with nonzero triality. More explicitly,
one has
\begin{multline}
\langle z_x z_y z_w \prod_v z_v^{n_v} \rangle_{3^{N_C}} = \delta_{C_x,C_y} \cdot \delta_{C_y,C_w} \cdot \prod_C W_0(C)\\
 + (1-\delta_{C_x,C_y}) \cdot  \delta_{C_y,C_w}\cdot W_2(C_x) \cdot W_1(C_y) \prod_{C\neq C_x,C_y} W_0(C)\\
 + (1-\delta_{C_x,C_y}) \cdot  \delta_{C_x,C_w}\cdot W_2(C_y) \cdot W_1(C_x) \prod_{C\neq C_x,C_y} W_0(C)\\
 + (1-\delta_{C_x,C_w}) \cdot  \delta_{C_x,C_y}\cdot W_2(C_w) \cdot W_1(C_x) \prod_{C\neq C_x,C_w} W_0(C)\\
 + (1-\delta_{C_x,C_y}) (1-\delta_{C_y,C_w}) (1-\delta_{C_x,C_w}) \cdot W_2(C_x) W_2(C_y) W_2(C_w) \prod_{C\neq C_x,C_y,C_w} W_0(C) 
\end{multline}
and we could now write down the expression for the partition function
$\llangle z_x z_y z_w\rrangle_{N_Q}$ in terms of $\delta$-functions
which simply single out the various contributions from the sectors
containing up to three clusters with nonzero triality. As before, in
the simulation we allow the breakup of a zero triality cluster with
$\sum_{x\in C} n_x>0$ into two clusters with triality $n_C = 1$ and 2,
respectively, as long as the resulting clusters contain the source
sites $x,y,w$ as given above. In addition, we also allow a further
breakup of the cluster with triality $n_C=1$ into two clusters with
triality $n_C=2$, so as to allow contributions of the kind when all
sources sit in different clusters.
 
By dividing $\llangle z_x z_y z_w \rrangle_{N_Q}$ by $Z_{N_Q}$ one
obtains an expression for the 3-quark potential in the background of
$N_Q$ quarks,
\begin{equation}
\exp(-\beta V_{q_x q_y q_w})|_{N_Q} = \frac{\llangle z_x z_y z_w \prod_v z_v^{n_v} \rrangle_{N_Q}}{\llangle \prod_v z_v^{n_v} \rrangle_{N_Q}} = \langle z_x z_y  z_w \rangle_{N_Q} \, .
\end{equation}
This expectation value can be measured by probing the extended
configuration space which allows for clusters with nonzero
triality. Denoting by $N_1$ and $N_2$ the number of clusters with
triality 1 and 2, respectively, the allowed configuration space is
restricted to configurations with $(N_1, N_2)=(0,0)$, $(1,1)$ and
$(0,3)$ nonzero-triality clusters. It is straightforward to generalize
the previous update steps in such a way that only configurations in
that restricted configuration space are generated.

Finally, we note that summing the 3-point quark correlator over all
positions $x,y,w$ yields the free energy difference between $N_Q = 0
\mod 3$ and $N_Q+3$ quarks, i.e.~the baryon free energy in the
background of $N_Q$ quarks,
\beq 
\exp(-\beta \Delta F_B)|_{N_Q} =
\exp(-\beta (F_B|_{N_Q+3} - F_B|_{N_Q})) = \frac{\llangle \prod_C
  \delta_{n_C,0}\rrangle_{N_Q+3}}{\llangle \prod_C
  \delta_{n_C,0}\rrangle_{N_Q}} = \frac{Z_{N_Q+3}}{Z_{N_Q}}\, .  
\eeq
This quantity gives a direct link between the baryon density $\rho_B =
N_Q/3V$ defined in the canonical formulation discussed here and the
quark chemical potential $\mu$ in the grand-canonical formulation
through the relation
\begin{equation}
\lim_{V\rightarrow \infty} \left. \frac{\beta}{3}
\Delta F_B \right|_{N_Q = 3\rho V}  = \frac{\beta}{3} \frac{\partial
  F_B}{\partial \rho_B} = \mu(\rho_B) \, .
\end{equation}

\section{Multiplicities of quark configurations and reweighting}
\label{sec:multiplicities}
The classification of the clusters and fermion number configurations
according to their triality guarantees the identification of those
configurations, for which the contributions cancel exactly in the
partition function or in the improved observables, so that we are left
with positive contributions only.  The positivity of configuration
weights and the existence of improved estimators provide a
complete solution of the complex action problem.

From a practical point of view, this is however not sufficient. While
it is easy to identify the relevant configurations, one also needs a
suitable algorithm to efficiently sample all the relevant
configurations. The algorithm discussed in the previous section
suffers for example from inefficiencies due to the fact that the
transition, e.g.~from the sector $(N_1, N_2) = (0,0)_{N_Q} \rightarrow
(0,1)_{N_Q-1}$ when simulating the ratio in eq.(\ref{eq:Delta F_q}),
is suppressed like $1/V$ at low density. Similar issues may arise in
the transitions $(0,0) \leftrightarrow (1,1)$ when calculating the
quark-antiquark correlation, or $(0,0) \leftrightarrow (1,1)
\leftrightarrow (0,3)$ in the calculation of the 3-point quark
correlator.  In this section we describe how these difficulties can be
circumvented by simulating only the $(N_1, N_2) = (0,0)_{N_Q}$ sector
with appropriate reweighting into all the other sectors relevant for
the various observables. The approach is based on the fact that the
bond configurations contributing to the partition function and the
improved observables are the same and the contributions only differ in
the multiplicities determined by the allowed quark number
configurations.
 
To be more precise, we recall the partition function in
eq.(\ref{eq:Z_NQ}) and note that the sum over the quark number
configurations for a fixed bond configuration is only affected by the
product of constraining $\delta$-functions. Hence, for a given cluster
configuration, which is controlled by the bond configuration $b$, we
define
\beq
\calN(N_Q,b) \equiv \sum_{\{n\}} \prod_C \delta_{n_C,0} 
\eeq
as the multiplicity of the bond configuration $b$, i.e.~the number of
quark configurations compatible with the constraints. The partition
function for $N_Q$ quarks then becomes
\beq
Z(N_Q) = \sum_{\{b\}} (e^\gamma-1)^{N_b} 3^{N_C} \cdot \calN(N_Q, b) \,.
\eeq
Similarly, the correlation function can be written as
\beq
\av{z_x z_y^*}_{N_Q} = \frac1{Z_{N_Q}}
\sum_{\{b\}} (e^\gamma-1)^{N_b} 3^{N_C} \cdot    
\calN_{xy}(N_Q, b) \,,
\eeq
where we introduced the number of quark configurations compatible with
the new constraints as
\beq
\calN_{xy}(N_Q, b)\equiv \sum_{\{n\}}
\left[\delta_{C_x,C_y} \delta_{n_{C_x},0} + (1-\delta_{C_x,C_y}) \delta_{n_{C_x},2}
\delta_{n_{C_y},1} \right] \prod_{{C \neq C_x}\atop{C\neq C_y}}
\delta_{n_C,0}\,.
\eeq
We then have
\beq
\av{z_x z_y^*}_{N_Q} = \frac{\sum_{\{b\}} (e^\gamma-1)^{N_b} 3^{N_C} \calN_{xy}(N_Q, b)}{\sum_{\{b\}} (e^\gamma-1)^{N_b} 3^{N_C}    
\calN(N_Q, b)} = \av{\frac{\calN_{xy}(N_Q, b)}{\calN(N_Q, b)}}_{N_Q} \,,
\eeq
where the latter average is over bond configurations $\{b\}$
distributed according to the Boltzmann factor
$P(b)\propto(e^\gamma-1)^{N_b} 3^{N_C} \calN(N_Q, b)$.

The multiplicity of a bond configuration, $\calN(N_Q, b)$, can be
computed in two steps: first determine all possible partitions $\{B\}$
of the total baryon number $N_B = N_Q/3$ over the clusters $C$, and
secondly for each of these partitions compute the number of possible
arrangements of $n$ quarks within each cluster of volume $|C|$,
denoted by the multiplicity $\P(n, |C|)$ in the following.  Note that
$\P$ depends only on the number of quarks and the volume of the
cluster, while the shape of the cluster is irrelevant.  The
multiplicity can be computed using the recurrence relation
\beq 
\P(n, v) = \sum_{k=0}^{n_\text{max}} \P(n-k, v-1) \,, 
\eeq
together with the recurrence stopping condition $\P(n,1)=1$ for $n\leq
n_\text{max}$ and $0$ otherwise.  The recurrence relation is justified
by the fact that the number of ways of distributing $n$ quarks on $v$
sites can be computed by separating a site out and then counting how
many configurations have no quark on that site, $\P(n, v-1)$, how many
have one quark on that site, $\P(n-1, v-1)$, etc., and then adding up
all these contributions.

To generate all partitions $\{B\}$ of $N_B$ baryons into $N_C$
clusters one can use the Algorithm~\ref{alg:1} given in Appendix
\ref{app:Baryon partitions}. Note that the partitions only depend on
the number of clusters and are not affected by the sizes of the
clusters. However, since the number of partitions grows quickly with
the number of clusters and the baryon number we will use stochastic
estimators, as detailed further below, in order to evaluate the sum
over the baryon partitions.

So for a given baryon partition $B = \{B_C\}$, where $B_C$ denotes the
number of baryons in the cluster $C$, the multiplicity, i.e.~the total
number of quark arrangements fulfilling the triality-0 constraint,
is simply the product of quark multiplicities over all clusters.
Denoting with $N_B$ the total number of baryons in partition $B$, that
is $N_B\equiv \sum_C B_C$, we generate the partitions $B$ such that
$N_B= N_Q/3$ and we finally have
\beq 
\calN(N_Q, b) = \sum_{\{B\}} \prod_C \P(3 B_C, |C|) \,,
\eeq 
the multiplicity of the bond configuration $b$, i.e.~the number of
{\em triality-0} quark partitions for that bond configuration.

For $\calN_{xy}(N_Q, b)$ we need partitions that are similar to the
ones for $\calN(N_Q, b)$.  When $C_x=C_y$ we have $\calN_{xy}(N_Q,
b)=\calN(N_Q, b)$.  For the case when $C_x\neq C_y$ the trialities of
$C_x$ and $C_y$ are $2$ and $1$, respectively.  A partition satisfying
this condition can be generated from a triality-0 partition by moving
a quark from $C_x$ to $C_y$. All {\em triality-broken} partitions with
$(N_1,N_2)=(1,1)$ and $n_{C_x}=1, n_{C_y}=2$ are generated from
triality-0 partitions that allow this move, namely the ones with
$\sum_{v\in C_x} n_v > 0$ and $\sum_{v \in C_y} n_v < |C_y| \cdot
n_\text{max}$.  Denoting the set of triality-0 partitions with
$\B_{0,0} \equiv \{B\}_{(N_1=0,N_2=0)}$ and the triality-broken ones
with $\B_{1,2} \equiv \{B\}_{(N_1=1,N_2=2)}$ we then have
\beq
\frac{\calN_{xy}(N_Q, b)}{\calN(N_Q, b)} = \F(C_x, C_y)\quad\text{with}\quad
\F(C', C'')\equiv\frac{\sum_{{\B}_{1,2}} \prod_C \P(M_C, |C|)}
{\sum_{{\B}_{0,0}} \prod_C \P(M_C, |C|)} \,
\eeq
where $M_C = \sum_{x\in C} n_x$ is the number of quarks in cluster
$C$.  Above we stressed that the fraction $\F$ does not depend on $x$
and $y$ directly, but rather on the clusters that these sites belong
to. As such, for a given bond configuration $b$ we can compute
$\F(C',C'')$ for all cluster combinations $C', C''$ and then use it to
collect statistics for correlators at all distances.
For moderate number of clusters and quarks the calculation can be
carried out by listing and summing over all relevant baryon
partitions. Unfortunately, the number of such partition grows quickly
as the number of quarks and clusters increases. It is therefore
advisable to use an estimator for the fraction function,
\beq
\label{eq:F(C,C') stochastic estimator}
\F(C', C'') = \av{
\frac{\P(M_{C'}-1, |C'|)\cdot \P(M_{C''}+1, |C''|)} {\P(M_{C'},
  |C'|) \cdot \P(M_{C''}, |C''|)} 
}_{{\B}_{0,0}} 
\eeq
where the average if over triality zero partitions $B\in{\B}_{0,0}$
distributed according to the Boltzmann factor $\prod_C \P(M_C,
|C|)$. This can be easily generated using baryon moves from cluster to
cluster, accepted or rejected based on a Metropolis process.  Note
that we assumed that the function $\P$ is zero when the quark number
is negative or exceeds saturation.

There are two other two-point functions of interest, namely $\av{z_x
  z_y}$ and $\av{z_x^* z_y^*}$ which are defined as
\beqs
\av{z_x z_y}_{N_Q} &\equiv \frac1{Z_{N_Q}} \sum_b (e^\gamma-1)^{N_b}
3^{N_C} \cdot \calN_{xy}^{(-2)}(N_Q, b) \,,
\\
\av{z_x^* z_y^*}_{N_Q} &\equiv \frac1{Z_{N_Q}} \sum_b
(e^\gamma-1)^{N_b} 3^{N_C} \cdot \calN_{xy}^{(+2)}(N_Q, b) \,,
\label{eq:quark-quark and antiquark-antiquark correlation function}
\eeqs
with
\beq
\calN_{xy}^{(\pm2)}(N_Q, b)\equiv \sum_{{\{n\}}\atop{|n|=N_Q\pm2}}
\left[\delta_{C_x,C_y} \delta_{n_{C_x},\pm2} + (1-\delta_{C_x,C_y}) \delta_{n_{C_x},\pm1}
\delta_{n_{C_y},\pm1} \right] \prod_{{C \neq C_x}\atop{C\neq C_y}}
\delta_{n_C,0}\,.
\eeq
The calculation proceeds along the same steps as with the
quark-antiquark correlator, with a modified estimator for the
multiplicity fraction
\beq
\F^{(\pm 2)}(C, C') = \av{
\frac{\P(M_{C}\pm1, |C|)\cdot \P(M_{C'}\pm1, |C'|)} {\P(M_{C},
  |C|) \cdot \P(M_{C'}, |C'|)} 
}_{{\B}_{0,0}} 
\eeq
where the average is again over triality zero partitions
$B\in{\B}_{0,0}$ as in eq.(\ref{eq:F(C,C') stochastic estimator}).

Other observables of interest are the 1-point averages $\av{z}$ and
$\av{z^*}$ defined by
\beqs
\av{z_x}_{N_Q} &\equiv \frac1{Z _{N_Q}} \sum_b (e^\gamma-1)^{N_b}
3^{N_C} \cdot \calN_{x}^{(-1)}(N_Q, b)
\,, \\
\av{z_x^*}_{N_Q} &\equiv \frac1{Z_{N_Q}} \sum_b (e^\gamma-1)^{N_b}
3^{N_C} \cdot \calN_{x}^{(+1)}(N_Q, b)
\,,
\eeqs
with
\beq
\calN_{x}^{(\pm1)}(N_Q, b)\equiv \sum_{{\{n\}}\atop{|n|=N_Q\pm1}}
\delta_{n_{C_x},\pm1}  \prod_{C\neq C_x} \delta_{n_C,0}\,.
\eeq
The corresponding multiplicity ratios are again calculated using the
estimators
\beq
\F^{(\pm 1)}(C) = \av{
\frac{\P(M_{C}\pm1, |C|) } {\P(M_{C}, |C|)} 
}_{{\B}_{0,0}} \,.
\eeq

Finally, the quark chemical potential $\mu$ is given by the partition function ratio
\beq
\label{eq:quark chemical potential}
\mu(N_Q+3/2) = -\frac13 \log \frac{Z_{N_Q+3}}{Z_{N_Q}} \,.
\eeq
This ratio can be computed using the multiplicity ratio average
\beq
\frac{Z_{N_Q+3}}{Z_{N_Q}} = 
\av{\frac{\calN(N_Q+3,b)}{\calN(N_Q,b)}}_{N_Q} \,
\eeq
where the average is over bond configurations $\{b\}$ distributed
according to the Boltzmann factor $P(b)\propto(e^\gamma-1)^{N_b}
3^{N_C} \calN(N_Q, b)$. For a given baryon partition $B$ with
$|B|=N_Q/3$ on a fixed bond configuration $\{b\}$ we can generate a
baryon partition $B'$ with $|B'|=(N_Q+3)/3$ by adding one baryon to
any of the clusters. However, this map is not one-to-one, because
several $B$ partitions can generate the same $B'$ partition by adding
one baryon. The exact number of such partitions $B$ is $N_0(B')$ which
is the number of clusters that have $B'_C\neq 0$. Therefore, we have
\beq
\calN(N_Q+3, b)=\sum_{B\in{\B}_{0,0}} \sum_C
\frac1{N_0(B^C)}\prod_{c'} \P(M_{C'}+3 \delta_{C,C'}, |C'|)
\,,
\eeq
where $B^C$ is the quark partition generated from $B$ by adding $3$
quarks to cluster $C$. Thus, the multiplicity ratio can be computed
using the estimator
\beq
\frac{\calN(N_Q+3, b)}{\calN(N_Q, b)} = \av{\sum_c \frac1{N_0(B^C)} 
\frac{\P(M_C+3, |C|)}{\P(M_C,|C|)} }_{{\B}_{0,0}} \,.
\eeq

\section{Severity of the sign problem}
\label{sec:Severity of the sign problem}
A measure for the severity of the sign problem is given by the
expectation value of the complex phase of the full Boltzmann weight
measured in the ensemble with the absolute value of the Boltzmann
weight, i.e.~the so-called real action ensemble. The latter is
particularly simple in the canonical formulation because the complex
fermionic contributions are just a pure phase,
\begin{equation}
\prod_x z_x^{n_x} = \exp(i \sum_x n_x \phi_x)
\end{equation}
where $\phi_x \in\{0,\pm 2\pi/3\}$ is the phase of the $\Z(3)$ spin at
site $x$. Indicating the average w.r.t.~the absolute value of the
measure by the subscript ${}_{|\, \cdot\, |}$, we have
\begin{equation}
\langle \exp(i \sum_x n_x \phi_x) \rangle_{|\, \cdot\, |, N_Q} =
\frac{\llangle  \prod_x z_x^{n_x} \rrangle_{N_Q}}{\llangle 1 \rrangle_{N_Q}}
\end{equation}
which we recognize as eq.(\ref{eq:complex phase vev}). It is easy to
construct an improved estimator for this quantity. Integrating out the
spins we realize that the denominator becomes
\begin{equation}
\llangle 1 \rrangle_{N_Q} = \sum_{\{b\}}\left(e^\gamma -
  1\right)^{N_b} \cdot 3^{N_C} \cdot \sum_{\{n\}} 1 
=  \llangle 1 \rrangle_{N_Q=0} \cdot \P(N_Q,V) 
\end{equation}
where $\P(N_Q,V)$, defined in the previous section, just counts the
number of (unrestricted) quark configurations with $N_Q$ quarks on a
lattice with $V$ sites. The numerator on the other hand yields
\begin{equation}
\llangle \prod_x z_x^{n_x}  \rrangle_{N_Q} = \sum_{\{b\}}\left(e^\gamma -
  1\right)^{N_b}  \cdot 3^{N_C}  \cdot \sum_{\{n\}} \prod_C \delta_{n_C,0}
= \sum_{\{b\}}\left(e^\gamma -
  1\right)^{N_b} \cdot 3^{N_C} \cdot \calN(N_Q,b) 
\end{equation}
where $\calN(N_Q,b)$ from the previous section counts the number of
quark configurations compatible with the constraints given by the bond
configuration $b$. Essentially, the ratio simply calculates the
average fraction of {\it allowed} to {\it all} fermion configurations,
\begin{equation}
\langle \exp(i \sum_x n_x \phi_x) \rangle_{|\, \cdot\, |, N_Q} =
\av{\frac{\calN(N_Q,b)}{\P(N_Q,V)}}_{N_Q=0} 
\end{equation}
where the average on the r.h.s.~is over all bond configurations at zero
density. However, this ratio is not likely to be well estimated by
generating bond configurations at $N_Q=0$, because if the system is in
a different phase for $N_Q \neq 0$ the corresponding bond
configurations are expected to be qualitatively different from those
at $N_Q=0$. In order to avoid this well-known overlap problem, one can
estimate the average sign from simulations at nonzero density. In that
case one measures
\begin{equation}
\frac{\llangle 1 \rrangle_{N_Q}}{\llangle  \prod_x z_x^{n_x}
  \rrangle_{N_Q}} = \langle \exp(-i \phi) \rangle_{|\, \cdot\, |, N_Q} =
\av{\frac{\P(N_Q,V)}{\calN(N_Q,b)}}_{N_Q} \, 
\end{equation}
where we introduced $\phi = \sum_x n_x \phi_x$.  Yet another way to
calculate the average sign starts from the observation that
\begin{equation}
\langle \exp(i \phi) \rangle_{|\, \cdot\, |, N_Q} =
\frac{Z_{N_Q}}{Z_{N_Q=0} \cdot \P(N_Q,V)} 
= \frac{Z_3}{Z_0} \cdot \frac{Z_6}{Z_3} \cdot \ldots \cdot
\frac{Z_{N_Q}}{Z_{N_Q-3}} \cdot \frac{1}{\P(N_Q,V)}
\end{equation}
which can be rewritten using $\ln Z_{N_Q+3}/Z_{N_Q}= -3
\mu(\frac{3}{2}+N_Q)$, leading to
\begin{equation}
\ln \langle \exp(i \phi) \rangle_{|\, \cdot\, |, N_Q} = 
-3 \sum_{k=0}^{N_Q/3-1} \mu(\frac{3}{2}+ 3k) -
\ln \P(N_Q,V) \, .
\label{eq:snake formula for average sign}
\end{equation}
\begin{figure}[t!]
\centering
\includegraphics[width=0.49\textwidth]{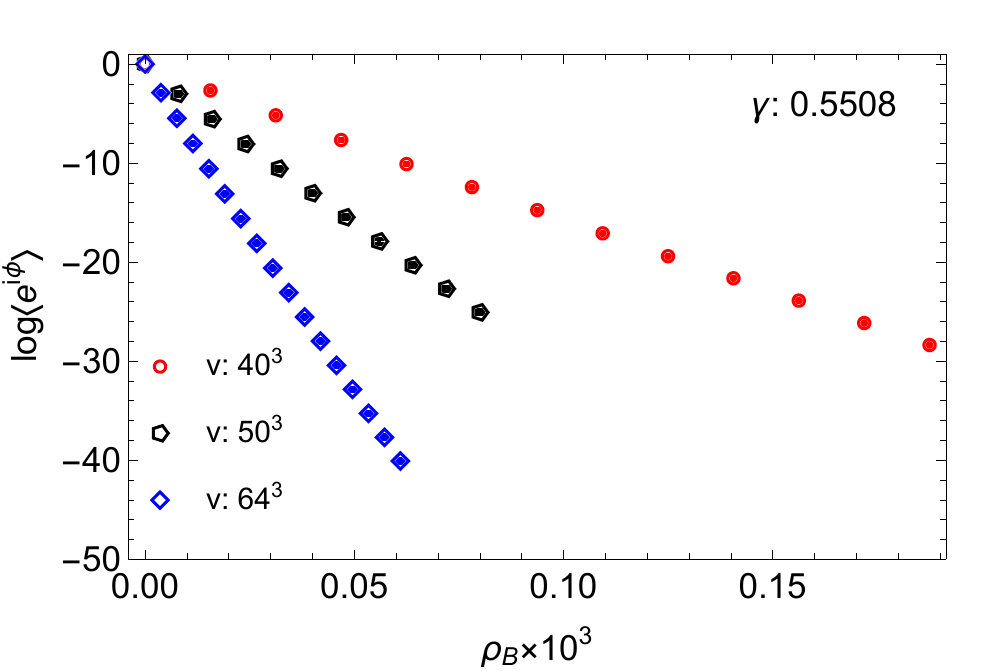}
\includegraphics[width=0.49\textwidth]{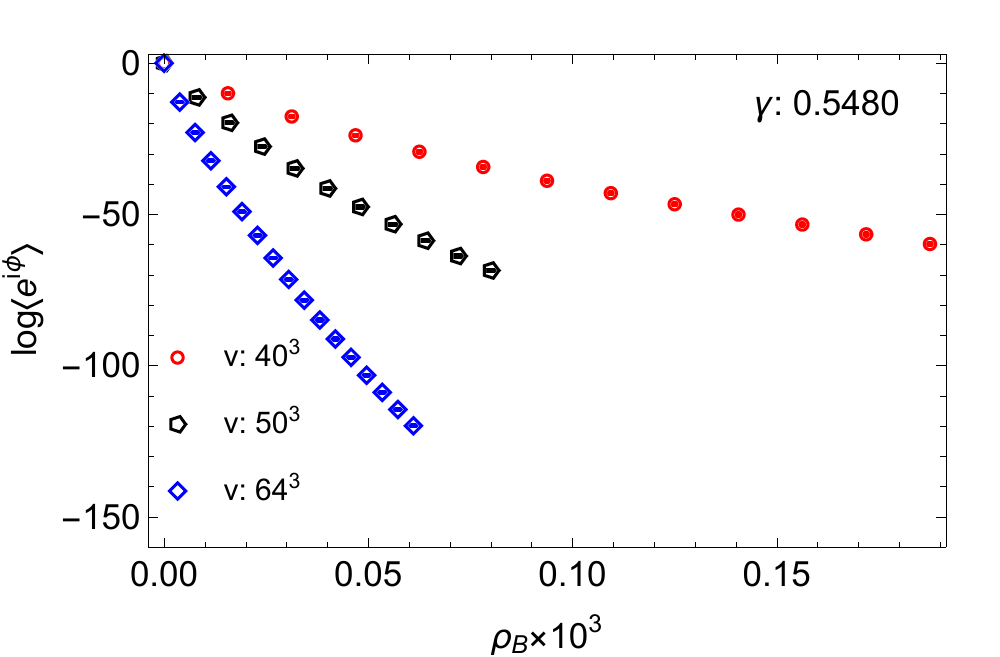}
\caption{The logarithm of the average phase $\ln \langle \exp(i \phi)
  \rangle_{|\, \cdot\, |}$ as a function of the baryon density
  $\rho_B$ for different volumes at high temperature in the deconfined phase ({\it left plot}) and at low temperature in the
  confined phase ({\it right plot}). 
  \label{fig:sign problem}}
\end{figure}

In Fig.~\ref{fig:sign problem} we show the logarithm of the average
phase $\ln \langle \exp(i \phi) \rangle_{|\, \cdot\, |}$, as obtained
through eq.(\ref{eq:snake formula for average sign}), as a function
of the baryon density $\rho_B$, defined by $\rho_B = (N_Q/3+1/2)/L^3$,
for different volumes and two different values of the coupling. The
left plot is for the coupling $\gamma=0.5508$ for which the system is
in the deconfined phase for all quark numbers, while the right plot is
for $\gamma=0.5480$ for which the system is in the confined phase for
all the quark numbers shown.  We observe that at fixed volume the
average sign goes to zero exponentially fast with the baryon density
with a rate that depends on the volume.

In order to quantify further the severity of the sign problem as it
scales with the volume at fixed baryon density, we define a scale
parameter $L_0$ which parameterizes the average sign as
$\exp(-V/L_0^3)$ at fixed density. In Fig.~\ref{fig:severity of sign
  problem} we show this scale parameter as a function of the baryon
density for three different couplings corresponding to three different
temperatures.
\begin{figure}[h!]
\centering
\includegraphics[width=0.75\textwidth]{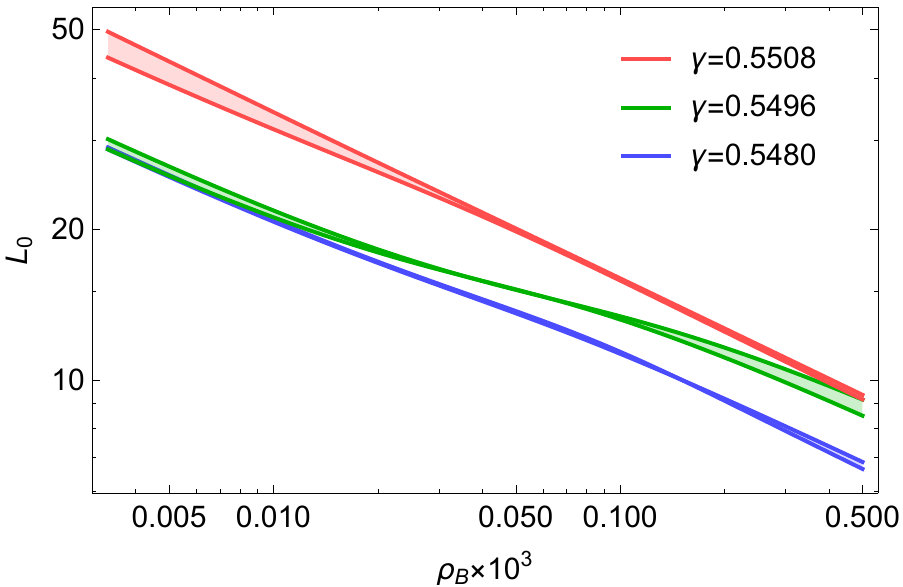}
\caption{The scale parameter $L_0$ related to the severity of the sign
  problem as a function of the baryon density for different
  couplings. The lower the value of $L_0$, the more severe the sign problem.
\label{fig:severity of sign problem}}
\end{figure}
The top curve corresponds to a high temperature for which the system
is in the deconfined phase for all densities, while the bottom curve
corresponds to a low temperature for which the system is confined for
the range of density values shown. The middle curve corresponds to an
intermediate temperature for which the system undergoes a first order
phase transition from the confined phase at low density into the
deconfined phase at higher density. This phase transition will be
discussed in more detail in the next section. Here we note that $L_0$
decreases linearly with the density, i.e.~the severity of the sign
problem gets worse exponentially with increasing density, independent
of whether the system is in the confined or deconfined phase. However,
the scale is distinctively different in the two phases and at fixed
density the sign problem in the confined phase is worse than in the
deconfined phase. The behaviour is consistent with the one observed in
the grand-canonical formulation (cf.~Fig.~4 in \cite{Alford:2001ug}),
but the sign problem appears to be more severe in the canonical
formulation.

\section{Results}
\label{sec:results}
From previous studies of the phase diagram in the Potts model
\cite{Karsch:2000xv,Alford:2001ug} one knows that the interesting
physics happens at very low density which is why we concentrated our
investigation of the sign problem  in the previous section on
that regime. We recall that the system
undergoes a first order deconfinement phase transition along a
critical line starting from the zero density transition point $(e^\mu,
\gamma) = (0, 0.550565(10))$ down to the critical endpoint at $(e^\mu,
\gamma) = (0.000470(2), 0.549463(13))$ where the system experiences a
second order phase transition in the universality class of the
3-dimensional Ising model.

\begin{figure}[t!]
\centering
\includegraphics[width=0.75\textwidth]{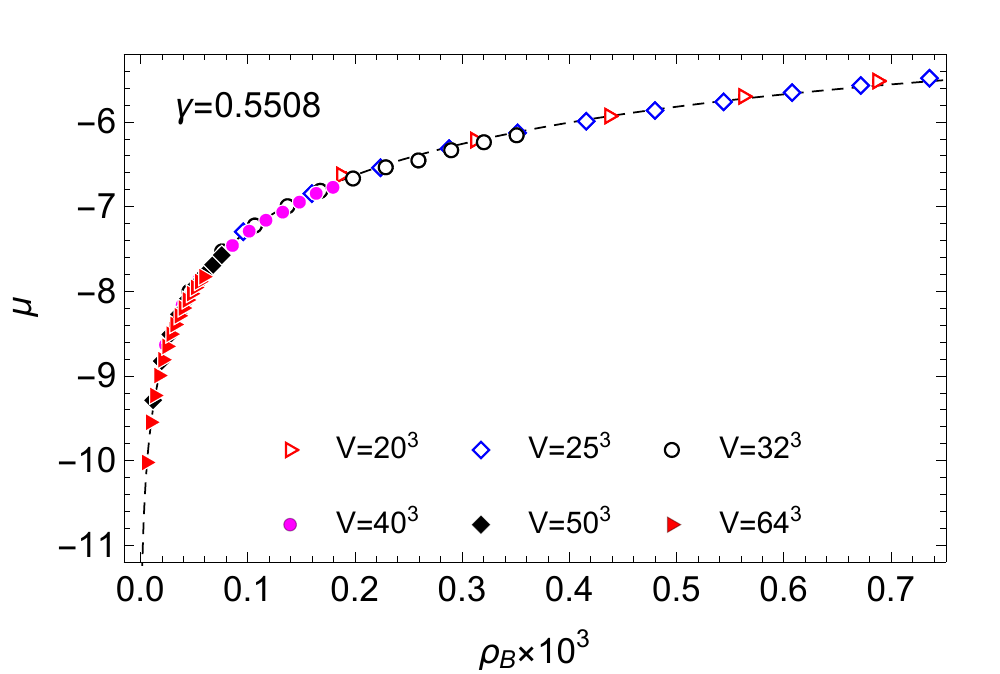}
\caption{The quark chemical potential $\mu$ as a function of the
  baryon density $\rho_B = (N_Q/3+1/2)/L^3$ in the deconfined phase at
  $\gamma=0.5508$ for various volumes $V=L^3$. The dashed line is a
  phenomenological, effective description of the function in terms of the behaviour of a (free) fermion system.
  \label{fig:rho_vs_mu_gamma0p5508_all}}
\end{figure}

We first discuss our results for the quark chemical potential $\mu$,
given by the partition function ratio in eq.(\ref{eq:quark chemical
  potential}), as a function of the baryon density. The function
essentially provides the translation between the canonical and
grand-canonical formulations. In
Fig.~\ref{fig:rho_vs_mu_gamma0p5508_all} we show the results for
various volumes $V=L^3$ at $\gamma=0.5508$ when the system is in the
deconfined phase.  For volumes $V \geq 25^3$ we observe practically no
finite volume effects and we find that the chemical potential grows
monotonically with increasing density. The functional dependence can
be well described by an ansatz for free fermions with an effective
fermion mass and we show the corresponding fit as the dashed line in
Fig.~\ref{fig:rho_vs_mu_gamma0p5508_all}.
\begin{figure}[b!]
\centering
\includegraphics[width=0.75\textwidth]{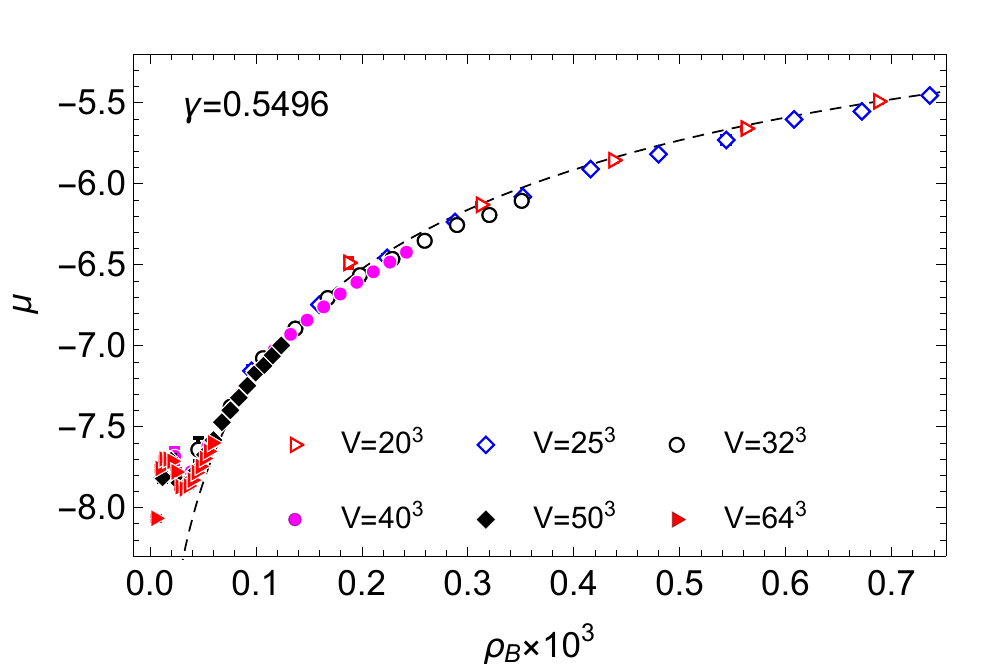}
\caption{The quark chemical potential $\mu$ as a function of the
  baryon density $\rho_B = (N_Q/3+1/2)/L^3$ for various volumes $V=L^3$ at
  $\gamma=0.5496$, just below the zero density
  deconfinement transition. The dashed line is a
  phenomenological, effective description of the function in terms
  of the behaviour of a (free) fermion system.
\label{fig:rho_vs_mu_gamma0p5496_all}}
\end{figure}
Next, in Fig.~\ref{fig:rho_vs_mu_gamma0p5496_all} we show the results
for the same quantity at $\gamma=0.5496$. At this coupling, the system
at zero density is in the deconfined phase, and our data clearly shows
a first order phase transition into the deconfined phase indicated by
the non-monotonic behaviour of the chemical potential $\mu$ as a
function of the density $\rho_B$.  The non-monotonicity is a typical
signature for a first order phase transition, and in
Fig.~\ref{fig:rho_vs_mu_gamma0p5496_zoom} we zoom in to inspect the
transition in more detail. The left plot illustrates how the behaviour
is better and better resolved towards the thermodynamic limit. Since
the transition happens at rather low density $\rho_B \lesssim 0.04
\cdot 10^{-3}$ one needs at least a lattice volume of $V=40^3$ to see
the onset of the non-monotonicity. The right plot shows the
determination of the critical chemical potential corresponding to the
first order phase transition on our largest volume $V=64^3$ using the
Maxwell construction.
\begin{figure}[t!]
\centering
\includegraphics[width=0.5\textwidth]{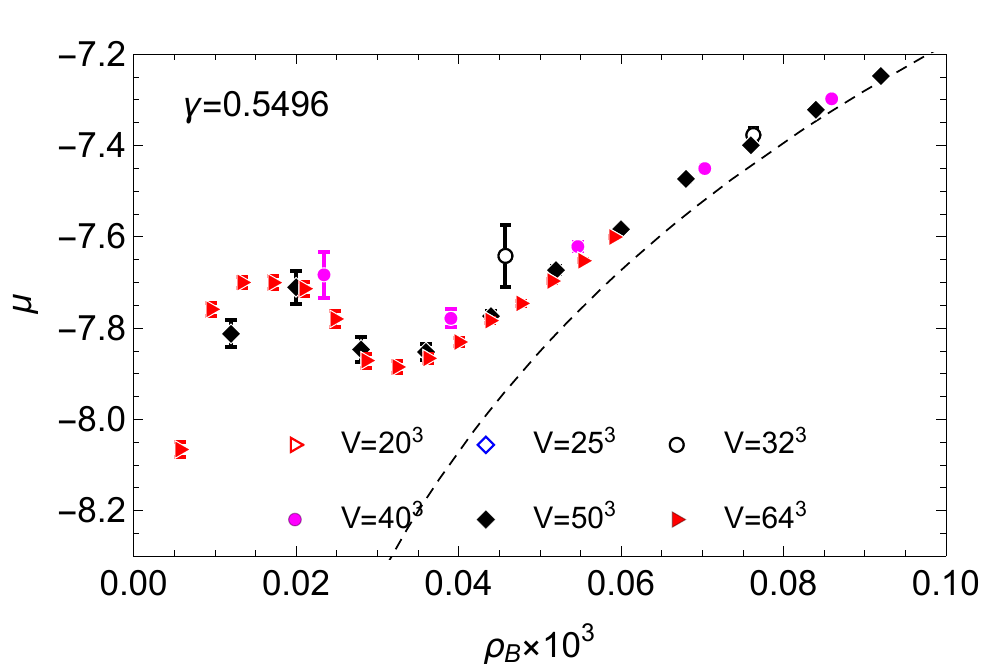}
\includegraphics[width=0.49\textwidth]{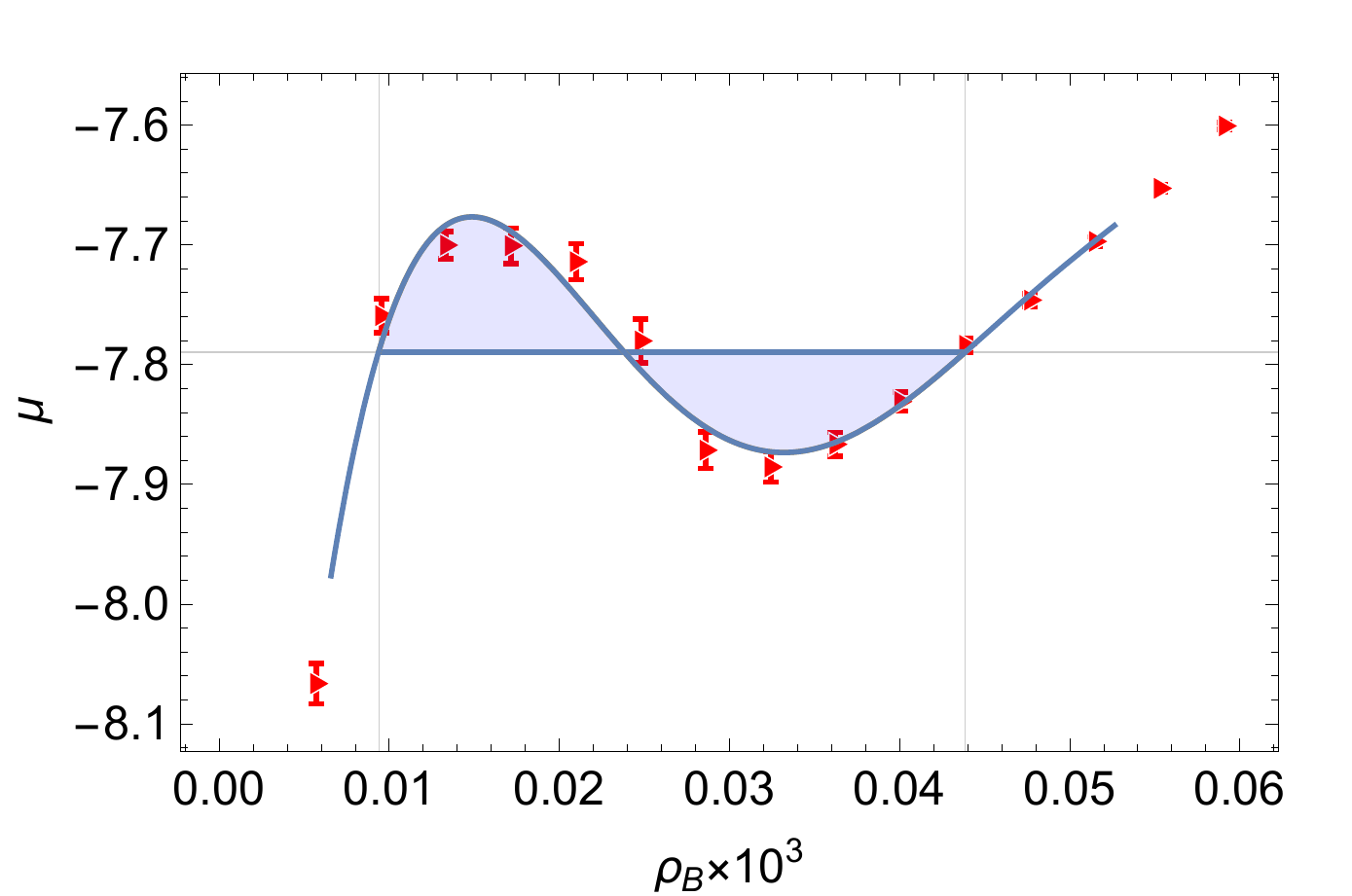}
\caption{The quark chemical potential $\mu$ as a function of the
  baryon density $\rho_B$ at $\gamma=0.5496$ zoomed into the density
  region where the phase transition occurs. {\it Left plot:} Finite
  volume behaviour. {\it Right plot:} Maxwell construction for 
  determining the critical chemical potential on the largest volume $V=64^3$. 
\label{fig:rho_vs_mu_gamma0p5496_zoom}}
\end{figure}
The construction guarantees that the two shaded regions limited by the
curve $\mu(\rho_B)$ and $\mu=\mu_c$ (as illustrated in the plot) have
the same area. The area can be related to the interface tension
associated with a first order phase transition.  From the Maxwell
construction one can also access the values of the upper and lower
densities $\rho_\text{up,low}$ at which the system enters or exits the
coexistence phase. A systematic investigation using various
interpolating functions and fit ranges yields $\mu_c=-7.7895(41)$ at
$\gamma=0.5496$ which agrees very well with the value determined in
\cite{Alford:2001ug}. For the upper and lower densities we obtain
$\rho_\text{up}=0.04386(53) \cdot 10^{-3}$ and $\rho_\text{low}=0.00940(14) \cdot 10^{-3}$.

\begin{figure}[t!]
\centering
\includegraphics[width=0.75\textwidth]{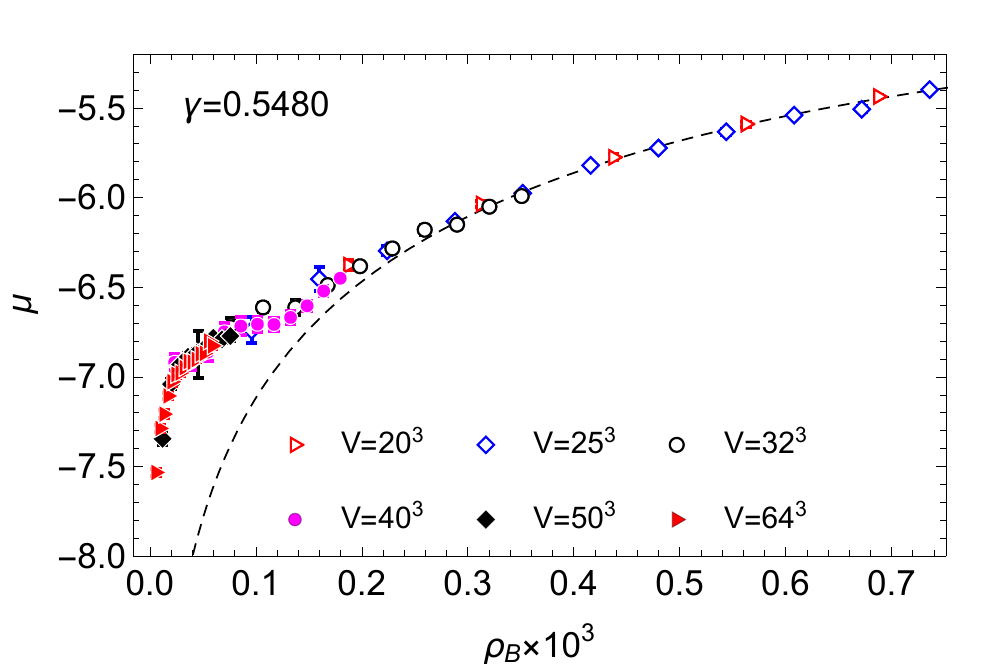}

\caption{The quark chemical potential $\mu$ as a function of the
  baryon density $\rho_B$ at $\gamma=0.5480$ for various volumes. A
  smooth cross-over transition from a (confined) baryon gas behaviour at low
  densities into a (deconfined) fermion gas behaviour at large
  densities is clearly visible.
\label{fig:rho_vs_mu_gamma0p5480_all}}
\end{figure}
In Fig.~\ref{fig:rho_vs_mu_gamma0p5480_all} we show once more the
quark chemical potential $\mu$ as a function of the baryon density for
various volumes, this time at $\gamma=0.5480$ which is well below the
critical endpoint. In this case, the system is in the confined phase
for small baryon densities and runs smoothly into the deconfined phase
at larger densities. The transition is no longer a phase transition,
but rather a smooth cross-over which presents itself as a monotonic
transition of the quark chemical potential from the baryonic gas form
to the one describing effectively free quarks. As in the previous
plots the dashed line is a phenomenological, effective description of
the behaviour in terms of a free fermion gas.

\begin{figure}[htb]
\centering
\includegraphics[width=0.49\textwidth]{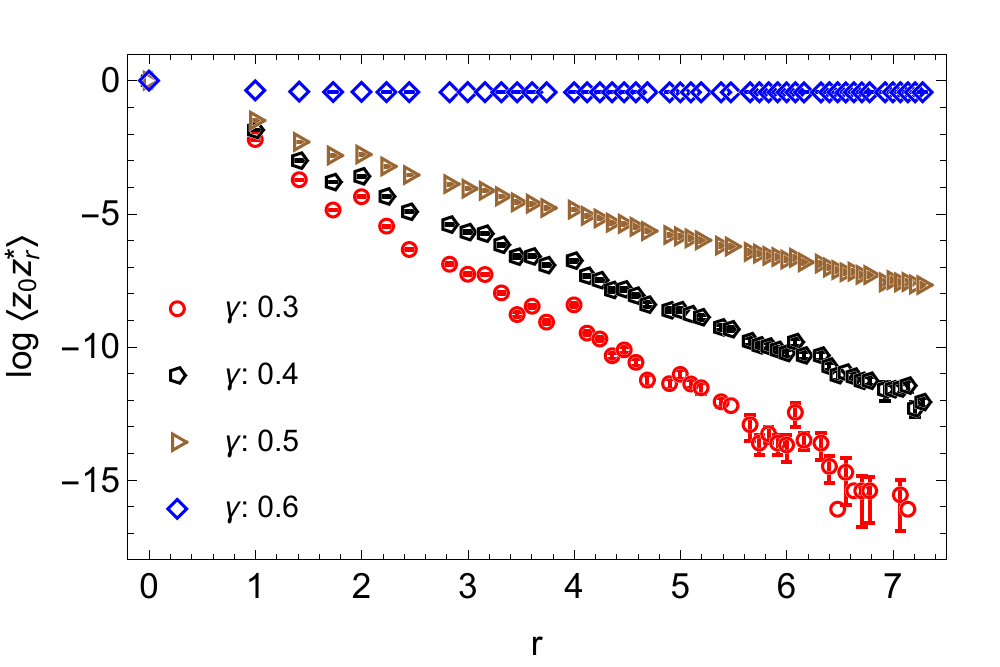}
\includegraphics[width=0.49\textwidth]{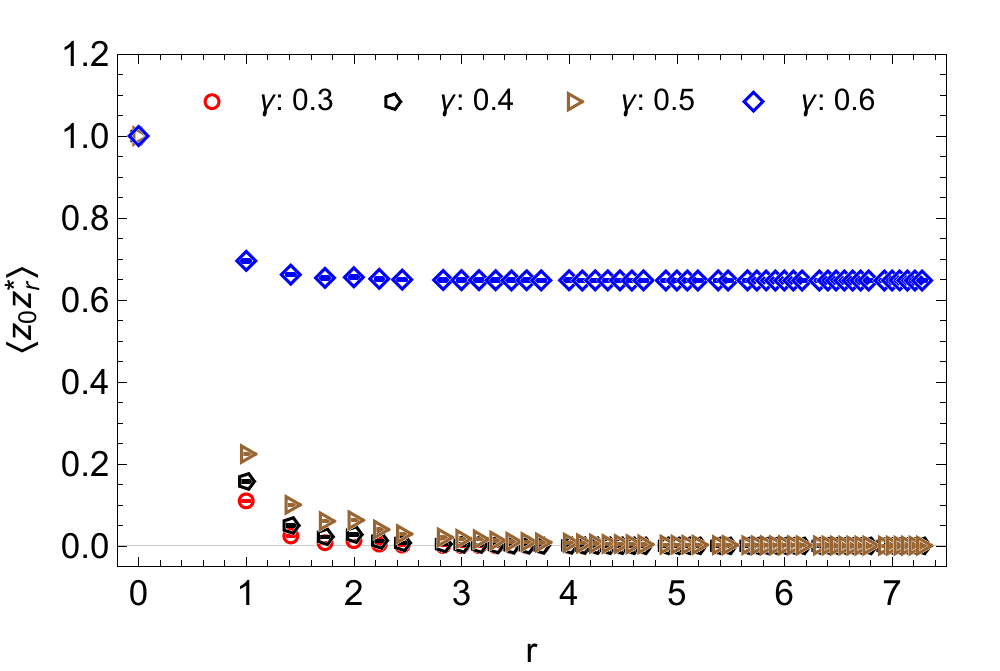}
\caption{The quark-antiquark correlator at zero density as
  a function of the separation $r$ in the confined phase at
  $\gamma=0.3, 0.4, 0.5$ and in
  the deconfined phase at $\gamma=0.6$ on a logarithmic scale ({\it
    left plot}) and on a linear scale ({\it right plot}). 
 \label{fig:potential_manygamma}}
\end{figure}

Next, we investigate the quark-antiquark correlator $\langle z_x
z_y^*\rangle_{N_Q=0}$ at zero density as a function
of the separation $r=x-y$ 
for three couplings $\gamma=0.3, 0.4, 0.5$ in the confined phase
 and for one coupling $\gamma=0.6$ in the deconfined phase 
in Fig.~\ref{fig:potential_manygamma}.
In the confined phase the correlator decays exponentially with the
distance for large distances, with a characteristic scale that becomes
longer for stronger coupling. This behaviour is consistent with the
fact that at zero density the free energies of a single quark and a
single antiquark are infinite because a single quark or antiquark can
not exist due to Gauss' law. This is reflected in the fact that the
expectation values for a quark or an antiquark being zero, or
equivalently the quark-antiquark correlator approaching zero at large
distances.  Above the phase transition point $\gamma=0.550565$, the
correlator approaches a constant at large distances signalling
deconfinement. This is illustrated in
Fig.~\ref{fig:potential_manygamma} by the data at $\gamma=0.6$.  The
finite asympotic value of the correlator at large distance indicates
the spontaneous breaking of the global $\Z(3)$ symmetry and reflects
the fact that the expectation values for a single quark and for a
single antiquark are nonzero in the deconfined phase.

\begin{figure}[t!]
\centering
\includegraphics[width=0.49\textwidth]{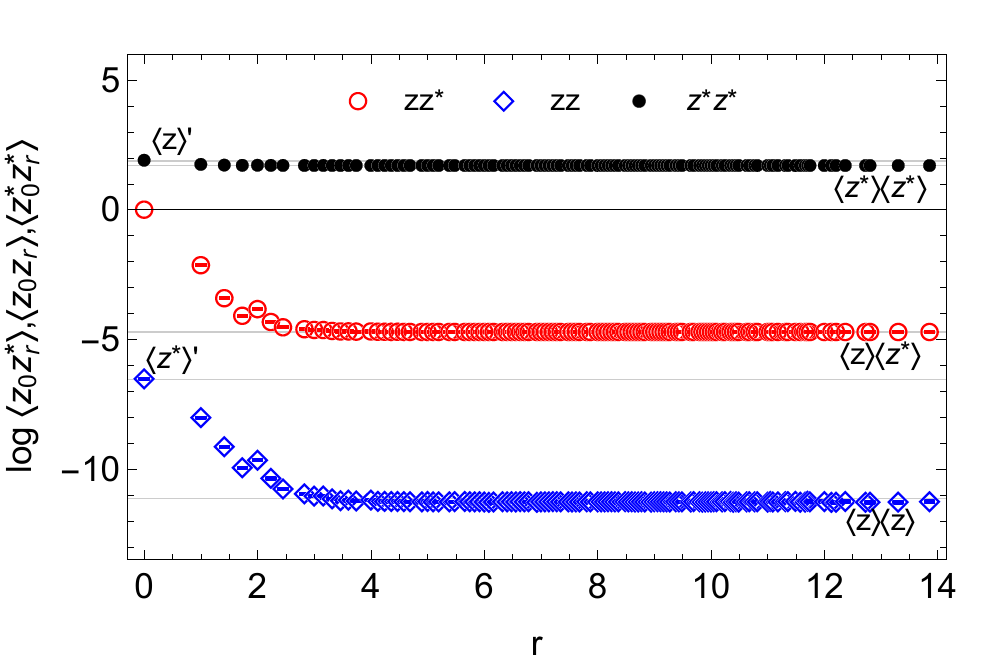}
\includegraphics[width=0.49\textwidth]{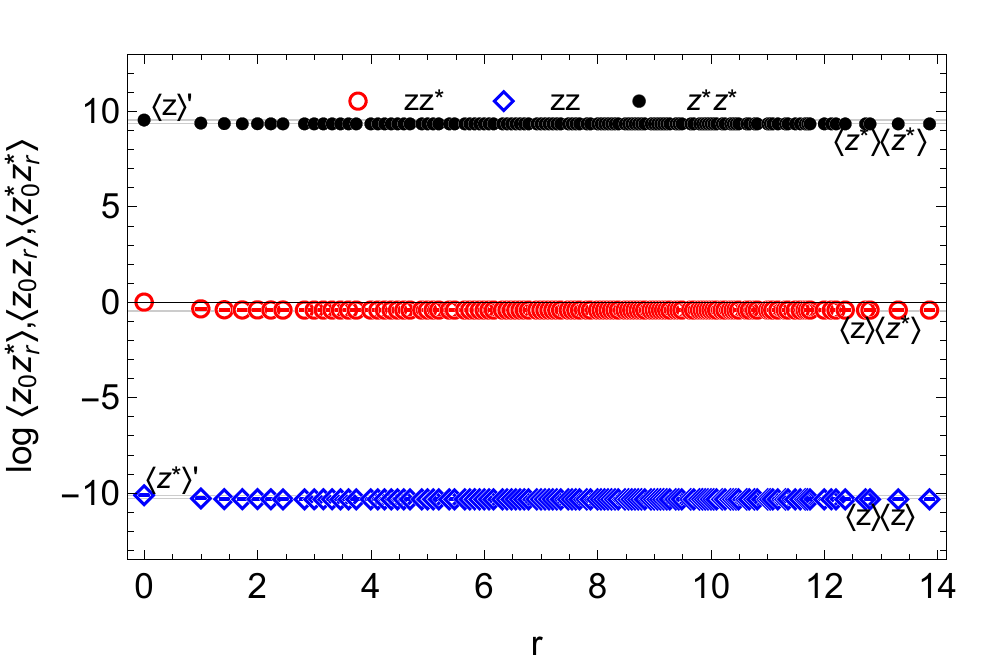}

\caption{The (anti)quark-(anti)quark correlators as a function
  of the separation $r$ in the confined phase at $\gamma=0.3$ ({\it left plot}) and in the deconfined phase at $\gamma=0.6$ ({\it right
  plot}) in a background of 8 baryons on a $V=16^3$ lattice, corresponding to the density
  $\rho_B \simeq 2.0 \cdot 10^{-3}$.
  \label{fig:all_potentials_gamma0p3_vol16_nq24_detail}}
\end{figure}

The situation is changed completely when the system is considered at
nonzero baryon density. Now, the free energies for a single quark and
a single antiquark are finite and the quark-quark, quark-antiquark and
the antiquark-antiquark correlators are screened at large
distance. This is illustrated in
Fig.~\ref{fig:all_potentials_gamma0p3_vol16_nq24_detail} where we show
the correlators as defined in eqs.(\ref{eq:quark-antiquark
  potential})-(\ref{eq:antiquark-antiquark potential}), as a function
of the separation $r=x-y$ for a system in the confined phase at
$\gamma=0.3$ (left plot) and in the deconfined phase at $\gamma=0.6$
(right plot), both in the background of 8 baryons on a $V=16^3$
lattice corresponding to a baryon density of $\rho_B \simeq 2.0 \cdot
10^{-3}$. For the following discussion it is useful to think of the
logarithm of the correlator as minus the potential energy (up to the
factor $\gamma$).

We first discuss the correlators in the confined phase (left plot). As
before, the quark-antiquark pair at distance $r=0$ annihilates and the
potential energy should therefore be zero. From eqs.(\ref{eq:improved
  estimator z zbar}) and (\ref{eq:quark-antiquark potential}) it is
clear that in our canonical cluster formulation this is exactly
realized with the improved estimator. At sufficiently large distances
the expectation value is expected to factorize into the expectation
values of a single quark and antiquark. The potential energy should
therefore flatten out and approach the sum of the free energies of the
corresponding single particles, in this case $F_q+F_{\bar q}$ obtained
from $\langle z\rangle \langle z^*\rangle$ as indicated in the figure
by the grey line. We note that our results indeed match the
expectation, not only qualitatively but also quantitatively.

For the quark-quark potential energy in the background of $N_B$
baryons we note that two quarks at the same position are equivalent to
having a single antiquark on top of an additional baryon, hence the
potential energy should be equal to the free energy of a single
antiquark in the corresponding background of $N_B+1$ baryons, obtained
from $\langle z^*\rangle'$. At large separation on the other hand, the
correlator is expected to factorize and the potential energy is given
as twice the free energy of a single quark, i.e.~$2 F_q$ obtained from
$\langle z\rangle \langle z\rangle$. Similarly, two antiquarks at the
same position are equivalent to a single quark on top of an
antibaryon, so the potential energy is equal to the free energy of a
single quark in the corresponding background with $N_B-1$ baryons
obtained from $\langle z\rangle'$, while at large separation it
approaches twice the free energy of a single antiquark, i.e.~$2
F_{\bar q}$ as obtained from $\langle z^*\rangle \langle
z^*\rangle$. In
Fig.~\ref{fig:all_potentials_gamma0p3_vol16_nq24_detail} these
energies, respectively the logarithms of the corresponding expectation
values, are again given as grey lines, and they illustrate that our
results qualitatively match the expectations. The small differences we
observe between the lines and the correlator data are finite volume
corrections.

We further note that the asymptotes of the correlators at large
distance are approximately equidistant. This is due to the fact that
in each step from bottom to top, a quark is replaced by an antiquark,
which approximately costs the same amount of free energy. One last
feature to point out is the flatness of the antiquark-antiquark
correlator as compared to the other correlators. It suggests that
screening two antiquarks is almost as effective as screening a single
quark in the background with one less baryon, i.e.~the system in the
confined phase screens antiquarks much more effective than
quarks. This is of course due to the fact that the system only
contains static quarks but not antiquarks as dynamical degrees of
freedom.

Let us finally discuss the correlators at finite density in the
deconfined phase at $\gamma=0.6$ (right plot). The discussions above
for the confined phase essentially carry over without change, the only
exception being that now all the correlators are very flat, i.e.~they
change by less than an order of magnitude over the whole range of
distances. This indicates that in the deconfined phase a quark and an
antiquark are screened equally efficiently, i.e.~the system does no
longer discriminate between a quark and an antiquark.

\section{Conclusions}
\label{sec:conclusions}
We have used a cluster algorithm to solve the notorious sign problem
in the Potts model approximation of heavy-dense QCD, i.e.~QCD with
heavy quarks and large chemical potential. Our approach is based on
the fact that in the canonical formulation the position of the quarks
determine the triality of the clusters, i.e.~the transformation
properties of the weights of the clusters under $\Z(3)$
transformations. This in turn allows to identify those configurations
for which the contributions to the partition function cancel exactly
after averaging each cluster over its $\Z(3)$ orientations, such that
only configurations remain which yield positive contributions. In this
sense, the concept of cluster triality is similar to the concept of
meron clusters formulated in different contexts
\cite{Chandrasekharan:1999cm} and hence the algorithm formulated here
belongs to the class of meron-cluster algorithms
\cite{Chandrasekharan:2002vk,Widmer:2011}.

The factorization of the weights into clusters with the subsequent
subaveraging of each cluster immediately yields an increase of
statistics by a factor $3^{N_C}$ where $N_C$ is the number of clusters
in a configuration. Since the number of clusters grows with the
volume, at otherwise fixed parameters, the gain in statistics is
exponential in the volume. We note that the mechanism at work here is
similar to the one formulated in \cite{Bloch:2013ara,Bloch:2015iha}
using the subset method. Here it is made to work for a
three-dimensional system and away from the strong coupling limit.
Furthermore, the triality properties of the clusters also allow the
construction of improved estimators which receive only positive
contributions. This is of course crucial for the construction of
efficient simulation algorithms.

We note that the solution of the sign problem for the Potts model
presented here is not the first one. In \cite{Alford:2001ug} the
authors constructed a cluster algorithm for the Potts model at finite
density in the grand-canonical formulation which solved the sign
problem completely. Another approach is to use the density of states
method \cite{Gattringer:2015lra}. The sign problem can also be avoided
by formulating the Potts model in the flux representation
\cite{Patel:1983sc,DeGrand:1983fk,Condella:1999bk,Alford:2001ug,Kim:2005ck,Mercado:2011ua,Mercado:2012yf}.
For this reason the focus of the present paper was not so much on the
physics but rather on the mechanism which forms the basis for the
solution of the sign problem. In that sense the solution here can be
considered as a proof of a concept which can be applied in other
cases.

Various extension of our solution of the sign problem in the canonical
formulation are possible. In this paper, following the spirit of
\cite{Alford:2001ug} we have only considered positive quark occupation
numbers, corresponding to a system with only quarks but no
antiquarks. In the context of QCD, a more natural scenario would be to
also allow for antiquarks which can form antibaryons and mesons. The
triality constraints for the contributions to the partition functions
and improved estimators carry over to this case without
modifications. The fermion number update described in
Sec.~\ref{sec:bond formulation} and \ref{sec:improved estimators}
becomes even simpler, because a quark-antiquark pair can be generated
within a cluster using the update step in eq.(\ref{eq:no merons
  fermion hop}) where the first constraint is replaced by
$(1-\delta_{n_x,-n_\text{max}})$. The possibility of this step
enhances the efficiency of the update considerably, especially at low
densities. On the other hand, the reweighting with the multiplicities
as described in Sec.~\ref{sec:multiplicities} becomes more complicated
because the calculation of the multiplicities is more complex.

Applying our findings in the canonical framework to other systems at
finite density is of course highly desirable. As an example we
mention QCD at fixed baryon number, because this is relevant in
heavy-ion collisions, and for studying few nucleon systems at low temperature. 
The application to QCD requires the extension of our approach to the gauge group
SU(3).
In this case one would
apply the method only to the discrete center of the group, but it is
conceivable that the discrete subset averaging is sufficient to solve
or greatly ameliorate the sign problem in QCD
\cite{Bloch:2013ara,Bloch:2015iha}. In order for this to work, one
needs to embed cluster variables in continuous groups
\cite{Niedermayer:1988kv,Niedermayer:1996ea}, in particular $\Z(3)$
into SU(3) \cite{Onofri:1992ke}. In \cite{BenAv:1991ve} a cluster
algorithm has been constructed along these lines for SU(2) on a single
time slice. Extending this construction to SU(3) and combining it with
the approach outlined in this paper is currently under investigation.

\section*{Acknowledgments}
We would like to thank Dan Boss, Uwe-Jens Wiese and Andreas Wipf for
useful discussions. A.A.~is supported in part by the National Science Foundation CAREER grant PHY-1151648 and by U.S.~DOE Grant No.~DE-FG02-95ER40907. G.B.~acknowledges support from the Deutsche
Forschungsgemeinschaft (DFG) Grant No.~BE 5942/2-1.

\begin{appendix}

\section{Connection between the grand-canonical and the canonical
  partition functions}
\label{app:connection grand-canonical/canonical}
While the grand-canonical and canonical partition functions are in
general equivalent and yield the same physics in the thermodynamic
limit, the exact correspondence depends on the details such as the
maximal fermion occupation number $n_\text{max}$ and the specific form
of the canonical fermion weights $g[z,n]$. Here we establish the
explicit relation between the canonical partition function in
eq.(\ref{eq:canonical partition function}), with the fermionic weights
as given in eq.(\ref{eq:simple fermionic weight}), and the
corresponding grand-canonical partition function. Since the
grand-canonical partition function in eq.(\ref{eq:grand-canonical
  partition function}) does not account for the fermionic nature of
the Potts spins at large local densities, the correspondence between
the grand-canonical partition function in eq.(\ref{eq:grand-canonical
  partition function}) and the canonical one in eq.(\ref{eq:canonical
  partition function}) is exact only in the low-density regime where
$e^\mu \equiv h \ll 1$, as we show below.

Using the fugacity expansion
\beq
Z_\text{GC}(h) = \sum_{N_Q=0}^\infty h^{N_Q}
Z_\text{C}(N_Q)
\eeq
we find that the associated grand-canonical partition function is
\beqs
Z_\text{GC}(h) = &\int{\cal D}z\,e^{-S[z]}
\prod_x \sum_{n_x=0}^{n_\text{max}} (z_x h)^{n_x}
=\int{\cal D}z\,e^{-S[z]}
\prod_x \frac{1-(z_x h)^{n_\text{max}+1}}{1-z_x h} \\
=&\int{\cal D}z\,e^{-S[z] -\sum_x\log[1-z_xh]
+\sum_x \log \left[1-(z_x h)^{n_\text{max}+1} \right] } \,.
\eeqs
To derive the relation above we need that $h<1$
which is the case for the simulations discussed in this paper. 
We note that $z_x^3=1$ and $n_\text{max}=0 \mod 3$. Using the
identity
\beq
\log[1-z_x f] = \frac{1}{3}\log[1-f^3] 
- z_x f \cdot {}_2F_1( \frac{1}{3} ; 1 ; \frac43 ; f^3)
- \frac12 z_x^* f^2\cdot {}_2F_1(\frac23; 1 ; \frac53 ; f^3) \,,
\eeq
which is valid for $|f|<1$, we find that the effective action for the
grand-canonical partition function is then
\beqs
S_\text{eff}=&S[z] + \sum_x z_x \left[ 
h \cdot {}_2F_1(\frac13 ; 1 ; \frac43 ; h^{3}) - 
h^{(n_\text{max}+1)}h \cdot  {}_2F_1(\frac13 ; 1 ; \frac43 ; h^{3(n_\text{max}+1)}) 
\right]\\
&+\sum_x z_x^* \frac12 
\left[ 
h^2 \cdot {}_2F_1(\frac23 ; 1 ; \frac53 ; h^{3}) - 
h^{2(n_\text{max}+1)} h \cdot {}_2F_1(\frac23 ; 1 ; \frac53 ; h^{3(n_\text{max}+1)}) 
\right] \,.
\eeqs
For our simulations we have $h$ of the order of $10^{-3}$ so the
higher order terms, ${\cal O}(h^{2})$, contribute very little. If we
keep only the first order terms our effective action simplifies to
\beq
S_\text{eff}=S[z]+ h \sum_x z_x + {\cal O}(h^{2})\,.
\eeq
This is the same action as the one in the Potts model studied by
Alford~et~al.~in \cite{Alford:2001ug}, hence we are able to reproduce
the phase diagram at low density, including the transition point at
zero density $(h, \gamma)=(0, 0.550565(10))$ and the critical endpoint
at $(h, \gamma)=(0.000470(2),0.549463(13))$, described in Fig.~1 of
that paper. Note that in \cite{Alford:2001ug} the parameter $\gamma$
is denoted as $\kappa$.

\section{Fully-dynamic connectivity problem}
\label{app:fully-dynamic connectivity}
The Potts model in the bond formulation, as given by the partition
function in eq.(\ref{eq:Swendsen-Wang cluster partition function}),
belongs to the class of random cluster models. Simulating such systems
with a local update algorithm requires one to check whether the bond
under consideration is a bridge or not. A bridge bond connects two
otherwise separate clusters when activated, or equivalently, breaks a
cluster into two disjoint clusters when deactivated. The determination
of the dynamically changing connectivity of bond clusters is a highly
complex problem which in graph theory and computer science is well
known under the name {\it fully-dynamic connectivity problem}. The
issue here is to find data structures and algorithms with a minimal
computational complexity for connectivity queries for the clusters,
while dynamically adding and deleting bonds.

In the context of the Potts model \cite{potts_1952}, the problem has
first been addressed by Sweeny in \cite{Sweeny:1980pj}, based on the
equivalence of the Potts model and the random cluster model described
by Fortuin and Kasteleyn \cite{Fortuin:1971dw}. An alternative
approach was suggested later by Swendsen and Wang
\cite{Swendsen:1987ce} and is based on alternating updates of spin and
bond variables which require only incremental connectivity
information.  It is well known that the Swendsen-Wang cluster
construction can be dealt with efficiently with the Hoshen-Kopelman
algorithm \cite{Hoshen:1976zz}. This is essentially a Find-Union
algorithm acting on weighted tree data structures. Combining it with
path compression to minimize the depth of the trees yields an
efficient algorithm for which connectivity queries essentially have
constant cost independent of the system size $L$.

Applying the same strategy to the situation of dynamically changing
connectivity leads to {\it computational critical slowing down},
because whenever a bond is updated, one needs to check whether that
bond acts as a bridge. Doing so requires to scan a number of bonds
proportional to the typical size of the clusters. Close to the phase
transition the clusters grow with the volume according to some
critical exponent $\beta$ and the computational cost hence grows like
${\cal O}(L^{\beta})$ \cite{Weigel:2013laa}.
In contrast, a nearly-optimal solution for the fully-dynamic
connectivity problem makes use of a spanning forest of Euler tour
trees \cite{Thorup:2000:NFG:335305.335345,Holm:2001:PDF:502090.502095}
in order to maintain a data structure which allows for an efficient
checking of the bridge property. The runtime of the corresponding
algorithm only grows as ${\cal O}(\ln^2 L)$ and we make use of it in
our implementation. We note that further improvements are possible
\cite{DBLP:journals/corr/abs-1209-5608} yielding asymptotically to a
computational complexity ${\cal O}(\ln^2 L/\ln \ln L)$ for bond
updates and ${\cal O}(\ln L/\ln \ln L)$ for connectivity
queries. Additional improvements can be obtained by using randomized
data structures \cite{Thorup:2000:NFG:335305.335345}. However, one
should consider that all these elaborate data structures involve some
computational overhead and it is not clear a priori which of the
strategies is most efficient. The fermion update in our algorithm for
example only requires connectivity queries for which the Find-Union
ansatz is most efficient.
Finally we note that the efficiencies of some of the strategies
discussed above have been investigated, and when possible compared to
the one of the Swendsen-Wang algorithm, for the generic random cluster
model in two dimensions in
\cite{Weigel:2013laa,1742-6596-510-1-012013}.

\section{Baryon partitions}
\label{app:Baryon partitions}
Here we describe an algorithm which generates all partitions of $N_B$
baryons over $N_C$ clusters.
\begin{algorithm}[tbh]
\begin{algorithmic}
\Procedure{Partitions}{$N_B$,$N_C$} 
\State $B\gets \{N_B, 0, 0, \ldots \}$ \Comment{$B$ is an array of $N_C$ entries}
\Loop
\State Print $B$.
\State Find smallest $k$ such that $B_k\not= 0$.
\State If $k=N_C$ exit.
\State $B_1 \gets B_k-1$.
\State If $k>1$ then $B_k \gets 0$.
\State $B_{k+1} \gets B_{k+1}+1$.
\EndLoop
\EndProcedure
\end{algorithmic}
\caption{Pseudocode to generate and print all partitions of $N_B$ baryons into $N_C$ clusters.}\label{alg:1}
\end{algorithm}
The logic of the algorithm is to impose an ordering relation between
different partitions of $N_B$ into $N_C$ non-negative numbers, and
then generate all of them by starting from the ``smallest'' one,
$B=\{N_B,0,\ldots,0\}$, and then successively generating the next
smaller one until we get to the ``largest'',
$B=\{0,\ldots,0,N_B\}$. The order is the usual lexicographic order,
that is, for two partitions $B=\{B_1, \ldots, B_{N_C}\}$ and
$B'=\{B_1', \ldots, B_{N_C}'\}$, $B$'s order with respect to $B'$ is
determined by the order of $B_{N_C}$ relative to $B_{N_C}'$. If
$B_{N_C}=B_{N_C}'$ then $B_{N_C-1}$ and $B_{N_C-1}'$ are compared.
This continues until we find the maximal rank $k$ where $B_k$ is
different from $B_k'$, and their order controls the order of $B$ and
$B'$. Of course $B=B'$ if (and only if) $B_k=B_k'$ for all
$k\in\{1,\ldots,N_C\}$. Algorithm \ref{alg:1} provides the pseudocode
for this procedure.

\end{appendix}

\bibliography{qcd_canonical_formulation}
\bibliographystyle{JHEP}

\end{document}